\documentclass{iopart}
\usepackage{iopams}
\usepackage{setstack}
\usepackage[all]{xy}

\def\appendixnumparts{\addtocounter{equation}{1}%
     \setcounter{eqnval}{\value{equation}}%
     \setcounter{equation}{0}%
     \def\theequation{\ifnumbysec
     \Alph{section}.\arabic{eqnval}{\it\alph{equation}}%
     \else\arabic{eqnval}{\it\alph{equation}}\fi}}
 
\def\endappendixnumparts{\def\theequation{\ifnumbysec
     \Alph{section}.\arabic{equation}\else
     \arabic{equation}\fi}%
     \setcounter{equation}{\value{eqnval}}}

\begin{document}
\title{Moyal star product approach to the Bohr-Sommerfeld approximation}
\author{Matthew Cargo, Alfonso Gracia-Saz, R G Littlejohn, 
M W Reinsch and P de M Rios}
\address{Departments of Physics and Mathematics, University of
California, Berkeley, California 94720 USA}

\ead{robert@wigner.berkeley.edu}

\begin{abstract}

The Bohr-Sommerfeld approximation to the eigenvalues of a
one-dimensional quantum Hamiltonian is derived through order $\hbar^2$
(i.e., including the first correction term beyond the usual result) by
means of the Moyal star product.  The Hamiltonian need only have a
Weyl transform (or symbol) that is a power series in $\hbar$, starting
with $\hbar^0$, with a generic fixed point in phase space.  The
Hamiltonian is not restricted to the kinetic-plus-potential form.  The
method involves transforming the Hamiltonian to a normal form, in
which it becomes a function of the harmonic oscillator Hamiltonian.
Diagrammatic and other techniques with potential applications to other
normal form problems are presented for manipulating higher order terms
in the Moyal series.

\end{abstract}

\pacs{03.65.Sq, 02.40.Gh, 03.65.Vf, 02.20.Sv}


\section{Introduction}

In this article we use the Moyal star product to derive the
Bohr-Sommerfeld approximation for the eigenvalues of the bound states
of a one-dimensional Hamiltonian, including higher order corrections
in $\hbar$.  We begin by quoting the result,
	\begin{equation}
	E_n = \left.\left[ H(A) + \frac{\hbar^2}{48}
	\frac{d}{dA} \left( \frac{1}{\omega(A)}
	\left< \{H,H\}_2 \right>_\phi\right) + O(\hbar^4) \right]
	\right|_{A=(n+1/2)\hbar},
	\label{BS2ndorder}
	\end{equation}
which uses the following notation.  $E_n$ is the $n$-th eigenvalue of
the quantum Hamiltonian $\hat H$, which has Weyl transform (or
``symbol'') $H$.  The latter is treated as a classical Hamiltonian
with action-angle variables $(A,\phi)$, and is regarded as a function
of the action $A$.  The frequency of the classical motion is
$\omega(A) = dH/dA$, and the notation $\{H,H\}_2$ refers to the second
Moyal bracket, defined in (\ref{Moyalbracket2def}) below.  This Moyal
bracket is otherwise twice the Hessian determinant of the Hamiltonian,
	\begin{equation}
	\{H,H\}_2 = 2[ H_{,\,xx} H_{,\,pp} - (H_{,\,xp})^2)].
	\label{MoyalHessian}
	\end{equation}
The angle brackets $\langle \ldots \rangle_\phi$ represent an average
over the angle $\phi$.  This result is discussed further in
subsection~\ref{BSrule}.

This paper assumes some background in the Wigner-Weyl formalism and
the Moyal star product.  A sampling of references in this area
includes Weyl (1927), Wigner (1932), Groenewold (1946), Moyal (1949),
Berry (1977), Balazs and Jennings (1984), Hillery \etal\ (1984),
Littlejohn (1986), McDonald (1988), Estrada
\etal\ (1989) and Gracia-Bond\'\i a and V\'arilly (1995).    

Our strategy is to use the Wigner-Weyl symbol correspondence and the
series representation of the Moyal star product to transform a given
Hamiltonian into a function of the harmonic oscillator, at least in a
neighborhood in phase space of a fixed point of the classical
Hamiltonian.  Notable aspects of this calculation are the higher order
terms in the Bohr-Sommerfeld formula, the use of the Moyal star
product to achieve a normal form transformation, and the development
of diagrammatic and other techniques for handling higher order terms
in the Moyal series.

In one-dimensional problems with Hamiltonians of the
kinetic-plus-potential form, it is straightforward to extend standard
WKB methods to higher order in $\hbar$ to find corrections to the
usual Bohr-Sommerfeld formula.  Results may be found in Bender and
Orszag (1978).  Several other approaches to the calculation of higher
order terms have been proposed in the literature (Maslov and Fedoriuk,
1981; Voros, 1977, 1989; Kurchan \etal\ 1989).  Our approach is
characterized by the use of the Weyl symbol correspondence for the
representation of operators, and the Moyal star product for the
manipulation of those operators.  In a subsequent paper we shall
extend our methods to the case of multidimensional, integrable systems
(torus quantization).

Recently Colin de Verdi\`ere (2004) has presented another approach for
calculating higher order corrections to the one-dimensional
Bohr-Sommerfeld formula that is based on the Weyl symbol
correspondence and the Moyal star product.  Colin de Verdi\`ere
concentrates on the spectrum of the operator, which is manipulated by
representing traces of operators as integrals over phase space.  In
his approach there is no need to examine eigenfunctions.  From an
algorithmic or computational standpoint, his method is quite simple,
the simplest we have seen for calculating higher order terms
(certainly simpler than ours).  On the other hand, our approach does
provide explicit representations of the transformations necessary to
construct eigenfunctions.  Another difference is that our method can
be generalized to the multidimensional case, whereas we could not see
how to do that with the method of Colin de Verdi\`ere.  The
generalization of our method was not trivial, however, and it may be
that with more effort his could be generalized, too.

The use of the Weyl symbol correspondence for representing operators
means that one can handle a wider class of Hamiltonians than the
kinetic-plus-potential type (second order differential operators).
For example, integral as well as differential operators are allowed.
This is important in applications, such as in plasma physics, where
such operators arise, and also for the multidimensional generalization
where not all the operators of the commuting set need be second order
differential operators. 

More importantly, the use of the Weyl symbol correspondence means that
the calculations take place in phase space, not configuration space.
For example, there is no trouble with caustics or turning points.
(Phase space methods are not necessarily free of caustic difficulties,
but caustics make no appearance in our approach.)  For another
example, the usual (lowest order) Bohr-Sommerfeld formula has an
obvious invariance under arbitrary canonical transformations, since
the energy eigenvalues are expressed in terms of the area of the
classical orbits in phase space.  To make this statement precise,
however, one would have to say precisely what the classical
Hamiltonian corresponding to a given quantum Hamiltonian is, over a
wider class of operators than those of the kinetic-plus-potential type
(a form that is not invariant under canonical transformations).  The
Weyl symbol correspondence does this, and provides a formalism that is
covariant under linear canonical transformations.  Thus, in approaches
based on the Weyl symbol correspondence, the lowest order
Bohr-Sommerfeld energy eigenvalues are invariant under aribitrary
canonical transformations, and the next and all higher order terms in
the $\hbar$ series are invariant under linear canonical
transformations.

The basic idea of this paper arises from the usual, lowest order
Bohr-Sommerfeld formula, which states that the eigenvalues of a
quantum Hamiltonian are given approximately by setting
$A=(n+1/2)\hbar$ in the classical formula expressing the classical
Hamiltonian $H$ as a function of its action $A$, $H=f(A)$.  This
formula suggests that the quantum Hamiltonian is a function of a
quantum ``action operator,'' of which the classical formula is a
lowest order representation by means of symbols, and that the
eigenvalues of the action operator are $(n+1/2)\hbar$.  Since these
are also the eigenvalues of the harmonic oscillator (of unit
frequency), the suggestion is that the action operator is unitarily
equivalent to the harmonic oscillator Hamiltonian.  If this is so,
then the quantum Hamiltonian is unitarily equivalent to a function of
the harmonic oscillator.  In this paper we find that these suggestions
are borne out, and we are able to construct explicitly the unitary
operator (or at least the symbol of its generator), in the sense of a
formal series in $\hbar$, which transforms a given quantum Hamiltonian
into a function of the harmonic oscillator.  We only require that the
quantum Hamiltonian have a ``slowly varying'' (defined below) Weyl
symbol, and that the symbol have a generic extremum (fixed point) at
some point in phase space.  The classical analog of the unitary
transformation we construct is a canonical transformation that maps
the level sets of the classical Hamiltonian around the extremum (which
are topological circles) into exact circles about the origin.  The
latter, of course, are the level sets of the harmonic oscillator.

The transformed Hamiltonian can be regarded as a normal form, that is,
a standard or especially simple form achieved by means of a
transformation.  In this case, the normal form is a function of the
harmonic oscillator, achieved on the level of $\hbar$-series for the
Weyl symbol and brought about by unitary transformations.  The class
of Hamiltonians which can be brought into this normal form are those
whose symbol has certain properties, notably that of having a family
of level sets of circular topology surrounding a generic fixed point.
This is the most generic case for bound states in a one-dimensional
system, and therefore the one to examine first.  But one can imagine
that there are other normal forms that apply in other cases, for
example, if the fixed point is not generic perhaps a standard quartic
oscillator is a normal form.  Or if in a certain region of
phase space the symbol has a separatrix surrounding two islands of
equal area and crossing in one unstable fixed point, perhaps a
standard, symmetric double well oscillator is a normal form.
Certainly at lowest order in $\hbar$ the unitary transformation is
represented by a canonical transformation that preserves area in phase
space, so the separatrix of the Hamiltonian and that of the normal
form must enclose the same area if they are to be unitarily equivalent.
Whether this is enough to guarantee the equivalence of the operators
(that is, the equality of their symbols to all orders in $\hbar$) is an
open question, as far as we know.

Normal form transformations of operators or sets of operators, either
by star product methods or by Fourier integral operators, have been
around for some time, in fact we ourselves have used the star product
to diagonalize or block diagonalize systems of coupled wave equations
(Littlejohn and Flynn, 1991) and to study problems of mode conversion
(Littlejohn and Flynn, 1992, 1993).  Such methods are well suited to
the development of systematic perturbation methods for quantum
adiabatic systems, such as molecules in the Born-Oppenheimer
approximation (Littlejohn and Weigert, 1993).  Similar normal form
transformations for coupled wave equations have also been discussed by
Braam and Duistermaat (1995), although with the idea of using Fourier
integral operators to carry out the transformations instead of star
products.  The star product approach to normal form transformations
for coupled wave equations (WKB on vector bundles) was put on firmer
mathematical foundations and generalized in various ways by Emmrich
and Weinstein (1996) and Emmrich and R\"omer (1998).  More recently,
Colin de Verdi\`ere \etal\ (1999) have studied normal forms for mode
conversion (generalized Landau-Zener transitions) and Colin de
Verdi\'ere and Parisse (1999) have studied them for problems involving
separatrices.  We also expect normal form theory to have applications
in transition state theory (Creagh, 2004).

In recent years there has arisen the subject of deformation
quantization, which involves generalizations of the Moyal star product
to nontrivial phase spaces (symplectic or Poisson manifolds).  The
phase space ${\mathbb R}^2$ used in this paper, upon which the Weyl
symbol correspondence is based, is considered trivial.  The general
idea is to deform the commutative algebra of multiplication of
functions on the phase space into a noncommutative but associative
algebra, where $\hbar$ is the deformation parameter and where the new
multiplication rule is the generalized star product.  It is also
required that the order $\hbar$ term in the symbol representation of
the commutator be proportional to the Poisson bracket.  The new
algebra is then interpreted as an algebra of operators on a quantum
system, the ``quantized'' version of the classical phase space.  In
one approach, the star product is represented as a formal power series
in $\hbar$, a generalization of the Moyal formula, and one must work
out the terms of the series subject to the constraint of associativity
and the appearance of the Poisson bracket at first order.  Basic
references in this area include Bayen \etal\ (1978), Fedosov (1994)
and Kontsevich (2003).  These works show an interesting geometrical
structure associated with the higher order terms in the Moyal star
product, which has stimulated our interest in higher order terms in
the Bohr-Sommerfeld formula.  Our results may be relevant for attempts
to understand eigenfunctions of an operator geometrically as
noncommutative versions of tori.

It should be possible to read the main body of this paper, skipping
the appendices, to obtain an overview of our calculation.  The
appendices, however, are needed for the details, including notational
conventions.

\section{The setup}

Let $\hat H$ be a Hermitian operator (the ``Hamiltonian'') in a
one-dimensional quantum system, that is, $\hat H$ acts on wave
functions $\psi(x)$, $x\in {\mathbb R}$ (the Hilbert space is
$L^2({\mathbb R})$).  We uniformly use hats ($\hat{\phantom{H}}$) over
a letter to denote operators, whereas a letter without a hat
represents the Weyl transform (or Weyl ``symbol'') of the operator.
For example,
	\begin{equation}
	H(x,p) =\int ds \, e^{-ips/\hbar} \,
	\langle x+s/2 \vert {\hat H} \vert x-s/2 \rangle
	\label{Weylsymboldef}
	\end{equation}
and
	\begin{equation}
	{\hat H} = \int \frac{dx \, dx'\, dp}{2\pi\hbar}
	e^{ip(x' - x)/\hbar} \, 
	H\left( \frac{x+x'}{2},p\right) \, 
	\vert x \rangle \langle x' \vert
	\label{Weyltransforminversedef}
	\end{equation}
illustrate the Weyl transform and its inverse in the case of the
Hamiltonian.  We regard $H$ as the ``classical Hamiltonian,'' defined
on the phase space ${\mathbb R}^2$, with coordinates $(x,p)$.  We
denote these coordinates collectively by $z^\mu=(x,p)$, $\mu=1,2$.

We assume that the symbol $H$ has an expansion in $\hbar$ beginning
with the power $\hbar^0$,
	\begin{equation}
	H = H_0 + \hbar H_1 + \hbar^2 H_2 + \ldots,
	\label{slowlyvaryingdef}
	\end{equation}
where each $H_n$ is independent of $\hbar$.  An operator whose symbol
possesses this type of expansion will be called ``slowly varying''.
Not all operators are slowly varying; for example, the unitary
operator $\exp(-i{\hat H}t/\hbar)$ is not.  The leading term ($H_0$ in
the example above) of the symbol of a slowly varying operator will be
called the ``principal symbol.''

We assume $H$ is smooth and has a generic extremum (a fixed point) at
some point of phase space.  The fixed point need not be at $p=0$, nor
does $H$ need to be invariant under time-reversal ($p \to -p$).  An
extremum is considered generic if the Hessian matrix $H_{,\,\mu\nu}$ of
the Hamiltonian is nonsingular at the extremum.  Here and below we use
comma notation for derivatives, for example,
	\begin{equation}
	H_{,\,\mu\nu} = \frac{\partial^2 H}{\partial z^\mu \partial
	z^\nu}.
	\label{commanotationdef}
	\end{equation}
For example, the fixed point $(x,p)=(0,0)$ of the quartic oscillator
($V(x)=x^4$) is not generic, because the Hessian matrix has rank 1 at
the fixed point. 

It is convenient in what follows to assume that the extremum is a
minimum (always the case for kinetic-plus-potential Hamiltonians).  If
not, we replace $\hat H$ by $-{\hat H}$ at the beginning of the
calculation.

Radial equations (on which $x$ is the radial variable $r\ge 0$) are
excluded from our formalism, because the Weyl symbol correspondence is
not defined in the usual way on the half line, and because the
centrifugal potential is singular.  We believe the best way to handle
such problems within a formalism like the one presented in this paper
is by reduction from a problem on a higher dimensional configuration
space ${\mathbb R}^n$ under some symmetry, typically $SO(n)$.  Nor are
singular potentials such as the Coulomb potential covered by this
formalism, because singularities generally invalidate the Moyal star
product expansion in $\hbar$, itself an asymptotic expansion.  The
usual lowest order Bohr-Sommerfeld formula usually does give correct
answers for singular potentials, at least to leading order in $\hbar$,
but the structure of the higher order terms (which powers of $\hbar$
occur, whether the corrections can be represented by powers of $\hbar$
at all, etc.) presumably depend on the nature of the singularity.

In view of our assumptions, the classical Hamiltonian $H$ has level
sets in some neighborhood of the fixed point that are topological
circles.  We concentrate on this region of phase space, and ignore any
separatrices and changes in the topology of the level sets of $H$
which may be encountered further away from the fixed point. 

For convenience we perform a canonical scaling on the coordinates
$(x,p)$ (or operators $(\hat x, \hat p)$) to cause them both to have
units of $\hbox{\rm action}^{1/2}$.  For example, in the case of the
ordinary harmonic oscillator, we would write $x'=\sqrt{m\omega}\,x$, $p'
= p/\sqrt{m\omega}$, and then drop the primes.  

The strategy will be to perform a sequence of unitary operations
that transform the original Hamiltonian $\hat H$ into a new
Hamiltonian that is a function of the harmonic oscillator Hamiltonian,
at least in the ``microlocal'' sense of the symbols in the
neighborhood of the fixed point.  Since unitary transformations do not
change the spectrum of an operator, the new Hamiltonian will have the
same spectrum as the old one.  But since the new Hamiltonian is a
function of the harmonic oscillator Hamiltonian, its eigenvalues are
easy to compute.

The transformations will proceed in two stages.  In the first stage,
we perform a ``preparatory'' transformation that maps $\hat H$ into a
new Hamiltonian $\hat K$ that is a function of the harmonic oscillator
Hamiltonian at lowest order in $\hbar$.  We follow this by a
sequence of near-identity unitary transformations that transform
$\hat K$ into a new Hamiltonian $\hat M$ that is a function of the
harmonic oscillator Hamiltonian to all higher orders in $\hbar$, at
least formally.  Thus, the stages are 
	\begin{equation}
	{\hat H} \to {\hat K} \to {\hat M}.
	\label{HKMstages}
	\end{equation}	

What we mean by the harmonic oscillator Hamiltonian is really the
action of the harmonic oscillator, given in operator and symbol form by 
	\begin{equation}
	{\hat I} = \frac{1}{2}({\hat x}^2 + {\hat p}^2),
	\qquad
	I = \frac{1}{2}(x^2 + p^2).
	\label{Idef}
	\end{equation}
It turns out that an operator is a function $\hat I$ if and only
if its symbol is a function of $I$, as will be discussed more fully
below, although the two functions are not the same beyond lowest order
in $\hbar$.  Thus, to ensure that the transformed Hamiltonian is a
function of $\hat I$, we require that its symbol be a function of $I$.

\section{The preparatory transformation}

The preparatory transformation (the first arrow in (\ref{HKMstages}))
is the most difficult, because it is not a near-identity
transformation and cannot be handled by Lie algebraic (power series)
methods.  This transformation will transform $\hat H$ into another
Hamiltonian $\hat K$ whose symbol is a function of $I$ plus terms of
order $\hbar^2$ and higher.  Thus, the principal symbol of $\hat K$
will be a function of $I$.  The preparatory transformation only makes
the leading order term in the symbol of $\hat K$ a function of $I$,
not the higher order terms.

\subsection{Imbedding $\hat H$ and $\hat U$ in a family}

Let $\hat H$ be given.  Ultimately, we shall seek a unitary
transformation $\hat U$ such that the new Hamiltonian $\hat K$,
defined by
	\begin{equation}
	{\hat K} = {\hat U} {\hat H} {\hat U}^\dagger,
	\qquad
	{\hat H} = {\hat U}^\dagger {\hat K} {\hat U},
	\end{equation}
has a symbol $K$ that is a function of $I$ plus terms of order
$\hbar^2$ and higher.

For the moment, however, it is conceptually simpler to imagine that
$\hat H$ and $\hat U$ are given, and to seek a means based on Weyl
symbols of computing $\hat K$, without regard to the functional form of
$K$.  We do this by imbedding $\hat U$ in a family, $0\le \epsilon \le
1$, that is by assuming that there exists a smooth family of unitary
operators ${\hat U}_\epsilon$, such that
	\begin{equation}
	{\hat U}_\epsilon = \left\{ 
	\begin{array}{ll}
	1 & \mbox{if $\epsilon=0$,} \\
	{\hat U} & \mbox{if $\epsilon=1$.}
	\end{array}\right.
	\label{Uepsiloncases}
	\end{equation}
The family ${\hat U}_\epsilon$ can be seen as a path in the group of
unitary transformations that act on Hilbert space, connecting the
identity and the final $\hat U$.  We do not assume $\epsilon$ is
small, and we do not carry out any power series expansion in
$\epsilon$.  We imbed $\hat H$ in a similar family, defining
	\begin{equation}
	{\hat H}_\epsilon = {\hat U}_\epsilon^\dagger \,{\hat K} \,
	{\hat U}_\epsilon,
	\label{Hepsilondef}
	\end{equation}
so that
	\begin{equation}
	{\hat H}_\epsilon = \left\{ 
	\begin{array}{ll}
	{\hat K} & \mbox{if $\epsilon=0$,} \\
	{\hat H} & \mbox{if $\epsilon=1$.}
	\end{array}\right.
	\label{Hepsiloncases}
	\end{equation}
One might say that the $\epsilon$-evolution runs backwards, since
$\hat K$ evolves into $\hat H$ as $\epsilon$ goes from 0 to 1.
As always, Weyl symbols of the operators above are denoted without the
hat, for example, $U$, $K$, $U_\epsilon$, $H_\epsilon$.  There are
several operators, symbols and functions in this paper that depend on
$\epsilon$, the notation for which is summarized in Table~\ref{table}.
We shall be interested in calculating ${\hat H}_\epsilon$, from which
$\hat K$ follows by setting $\epsilon=0$.

\begin{table}

\caption{\label{table}Notation for operators, symbols and functions
depending on $\epsilon$.}

\begin{indented}
\item[]\begin{tabular}{@{}lllllll}
\br 
$\epsilon=0$ & $1$ & $K$ & $K_n$ & $\hbox{\rm Id}^\mu$ & $\theta$ & $I$ \\
any $\epsilon$ & $U_\epsilon$ & $H_\epsilon$ & $H_{n\epsilon}$ 
	& $Z^\mu_\epsilon$ & $\phi_\epsilon$ & $A_\epsilon$ \\
$\epsilon=1$ & $U$ & $H$ & $\delta_{n0}\,H$ & $Z^\mu$ & $\phi$ & $A$ \\
\br
\end{tabular}
\end{indented}
\end{table}

We obtain a differential equation for ${\hat H}_\epsilon$ by
differentiating (\ref{Hepsilondef}), which gives
	\begin{equation}
	\frac{d {\hat H}_\epsilon}{d \epsilon} = 
	\frac{i}{\hbar} [ {\hat G}_\epsilon, {\hat H}_\epsilon],
	\label{Hepsilonode}
	\end{equation}
where the Hermitian operator ${\hat G}_\epsilon$ (the ``generator'')
is defined by
	\begin{equation}
	{\hat G}_\epsilon= i\hbar\, {\hat U}^\dagger_\epsilon \,
	\frac{d{\hat U}_\epsilon}{d\epsilon} =
	-i\hbar \, \frac{d{\hat U}^\dagger_\epsilon}{d\epsilon}
	\, {\hat U}_\epsilon = {\hat G}^\dagger_\epsilon.
	\label{Gepsilondef}
	\end{equation}
We assume that ${\hat G}_\epsilon$ is slowly varying.  We shall solve
(\ref{Hepsilonode}) by converting operators to symbols and using the
Moyal product formula.  See \ref{Moyalnotation} for the Moyal star
product and the diagrammatic notation we shall use for the functions
and operations that arise from it.

\subsection{Differential equations for $H_\epsilon$ and $H_{n\epsilon}$}

We now transcribe (\ref{Hepsilonode}) to symbols and substitute
(\ref{Moyalcommutator}).  This gives a differential equation for the
symbol $H_\epsilon$,
	\begin{equation}
	\frac{d H_\epsilon}{d\epsilon} = -\{G_\epsilon,H_\epsilon\}
	+\frac{\hbar^2}{24} \{G_\epsilon,H_\epsilon\}_3 
	-\frac{\hbar^4}{1920}\{G_\epsilon,H_\epsilon\}_5 + \ldots,
	\label{Hsymbolode}
	\end{equation}
which is subject to the boundary condition $H_\epsilon = H$ at
$\epsilon=1$.  We express the solution of this equation in terms of a
set of new functions or symbols, $H_{0\epsilon}$, $H_{2\epsilon}$,
etc., which are required to satisfy the differential equations,
	\numparts
	\begin{eqnarray}
	\frac{dH_{0\epsilon}}{d\epsilon} - 
	\{H_{0\epsilon},G_\epsilon\} &=0, 
	\label{Hsub0epsilonode} \\
	\frac{dH_{2\epsilon}}{d\epsilon} - 
	\{H_{2\epsilon},G_\epsilon\} &= 
	\frac{1}{24} \{G_\epsilon,H_{0\epsilon}\}_3,
	\label{Hsub2epsilonode} \\
	\frac{dH_{4\epsilon}}{d\epsilon} - 
	\{H_{4\epsilon},G_\epsilon\} &= 
	\frac{1}{24}\{G_\epsilon,H_{2\epsilon}\}_3 -
	\frac{1}{1920}\{G_\epsilon,H_{0\epsilon}\}_5,
	\label{Hsub4epsilonode}
	\end{eqnarray}
	\endnumparts
etc., and the boundary condition $H_{n\epsilon}=\delta_{n0}\, H$ at
$\epsilon=1$.  Then we have
	\begin{equation}
	H_\epsilon = H_{0\epsilon} + \hbar^2 H_{2\epsilon} + \hbar^4
	H_{4\epsilon}+ \ldots.
	\label{Hsubnepsilondef}
	\end{equation}
This is not an expansion of $H_\epsilon$ in powers of $\hbar$ as in
(\ref{slowlyvaryingdef}), because the functions $H_{n\epsilon}$ are
themselves allowed to have a dependence on $\hbar$.  But each of these
is slowly varying, so that if the series (\ref{Hsubnepsilondef}) is
truncated, the order of the omitted term is given by the $\hbar$
coefficient.  Finally, we define $K_n = H_{n\epsilon}$ evaluated at
$\epsilon=0$ (see Table~\ref{table}), so that we have an expansion of
the symbol $K$ of $\hat K$,
	\begin{equation}
	K = K_0 + \hbar^2 K_2 + \hbar^4 K_4 + \ldots.
	\label{Ksubndef}
	\end{equation}

The solutions of (\ref{Hsub0epsilonode})--(\ref{Hsub4epsilonode}) can
be expressed in terms of a certain $\epsilon$-dependent, classical
canonical transformation, $z^{\prime \mu}(\epsilon) =
Z^\mu_\epsilon(z)$, where $z$ and $z'$ are the old and new variables,
and $Z^\mu_\epsilon$ is the transformation function.  The family of
canonical transformations $Z^\mu_\epsilon$ reduces to the identity at
$\epsilon=0$, while at $\epsilon=1$ we shall denote the transformation
simply by $Z^\mu$ (without the $\epsilon$).  See Table~\ref{table}.
The transformation $Z^\mu_\epsilon$ will be defined momentarily, but
it turns out that the left hand sides of
(\ref{Hsub0epsilonode})--(\ref{Hsub4epsilonode}) are convective
derivatives along the associated Hamiltonian flow.
Equation~(\ref{Hsub0epsilonode}) is a homogeneous equation for the
unknown $H_{0\epsilon}$, and the others are inhomogeneous equations
with driving terms determined by lower order solutions.  The structure
of the system is that of a Dyson expansion, in which the canonical
transformation $Z^\mu_\epsilon$ specifies a kind of interaction
representation.  The definition of $Z^\mu_\epsilon$ requires some
notational understandings that are presented in
\ref{functionnotation}.

\subsection{The canonical transformations $Z$ and $Z_\epsilon$}

The canonical transformation $Z^\mu_\epsilon$ is defined as the
solution of the functional differential equation,
	\begin{equation}
	\frac{d Z^\mu_\epsilon}{d\epsilon} =
	\{Z^\mu_\epsilon, G_\epsilon\},
	\label{Zsubepsilonode}
	\end{equation}
subject to the initial conditions, $Z^\mu_\epsilon = {\rm Id}^\mu$ at
$\epsilon=0$, and we define $Z^\mu = Z^\mu_\epsilon$ at $\epsilon=1$
(see Table~\ref{table}).  The functions $Z^\mu_\epsilon$ so defined
constitute a canonical transformation, for if we compute the
$\epsilon$-derivative of their Poisson brackets among themselves, we
find
	\begin{equation}
	\frac{d}{d\epsilon} \{Z^\mu_\epsilon, Z^\nu_\epsilon\}=
	\{\{Z^\mu_\epsilon,G_\epsilon\},Z^\nu_\epsilon\} +
	\{Z^\mu_\epsilon,\{Z^\nu_\epsilon,G_\epsilon\}\} =
	-\{G_\epsilon,\{Z^\mu_\epsilon,Z^\nu_\epsilon\}\},
	\label{ZPBode}
	\end{equation}
where we have used the Jacobi identity.  These are subject to the
initial conditions $\{Z^\mu_\epsilon, Z^\nu_\epsilon\} = J^{\mu\nu}$
at $\epsilon=0$.  But since $J^{\mu\nu} = \hbox{\rm const}$, the
initial conditions are the solution for all $\epsilon$, as shown by
direct substitution.

The canonical transformation $Z^\mu_\epsilon$ is not generated by
$G_\epsilon$ regarded as an $\epsilon$-dependent Hamiltonian function,
but rather by $G'_\epsilon= G_\epsilon \circ Z_\epsilon^{-1}$.  That
is, if we write $z^\mu(\epsilon) =Z^\mu_\epsilon(z_0)$ for the
solution of Hamilton's equations,
	\begin{equation}
	\frac{dz^\mu}{d\epsilon} = J^{\mu\nu} \,
	G'_{\epsilon,\nu}(z),
	\label{zepsilonode}
	\end{equation}
then the functions $Z^\mu_\epsilon$ satisfy
	\begin{equation}
	\frac{dZ^\mu_\epsilon}{d\epsilon}=
	J^{\mu\nu}\,G'_{\epsilon,\nu} \circ Z_\epsilon
	= \{{\rm Id}^\mu, G'_\epsilon\} \circ Z_\epsilon
	= \{Z^\mu_\epsilon,G_\epsilon\},
	\end{equation}
which agrees with (\ref{Zsubepsilonode}).  In the final step we have
used an important property of the Poisson bracket, namely, that if $A$
and $B$ are any two functions and $Z$ is a canonical transformation
(symplectic map), then
	\begin{equation}
	\{A,B\} \circ Z = \{ A\circ Z, B\circ Z\}.
	\label{PBcompositionrule}	
	\end{equation}

\subsection{Notation for $\epsilon$-derivatives}

The following notation will be useful for carrying out
differentiations and integrations in the interaction representation,
specified by composing a function with $Z^{-1}_\epsilon$.

For any function $F_\epsilon$ on phase space, possibly
$\epsilon$-dependent, we define
	\begin{equation}
	\frac{DF_\epsilon}{D\epsilon} = 
	\left[\frac{d}{d\epsilon}(F_\epsilon \circ
	Z^{-1}_\epsilon)\right]\circ Z_\epsilon,
	\label{DDepsilondef}
	\end{equation}
for a kind of derivative operator in the interaction representation.
This can be written in an alternative form,
	\begin{equation}
	\frac{DF_\epsilon}{D\epsilon} = 
	\frac{dF_\epsilon}{d\epsilon} -
	\{F_\epsilon,G_\epsilon\}.
	\label{altDDepsilondef}
	\end{equation}
The proof of (\ref{altDDepsilondef}) is obtained by setting
$F'_\epsilon = F_\epsilon \circ Z^{-1}_\epsilon$, so that
	\begin{equation}
	\frac{dF_\epsilon}{d\epsilon} =
	\frac{d}{d\epsilon}(F'_\epsilon \circ Z_\epsilon)
	= \frac{dF'_\epsilon}{d\epsilon} \circ Z_\epsilon
	+ (F'_{\epsilon,\mu} \circ Z_\epsilon) 
	\frac{dZ^\mu_\epsilon}{d\epsilon}. 
	\end{equation}
But by (\ref{Zsubepsilonode}) and the chain rule for the Poisson
bracket, the final term can be written,
	\begin{equation}
	(F'_{\epsilon,\mu} \circ Z_\epsilon) 
	\{Z^\mu_\epsilon,G_\epsilon\} =
	\{F'_\epsilon \circ Z_\epsilon, G_\epsilon\} =
	\{F_\epsilon, G_\epsilon\}.
	\end{equation}
Rearranging the result gives (\ref{altDDepsilondef}).  

\subsection{Solutions for $H_{n\epsilon}$ and $K_n$}

In view of (\ref{altDDepsilondef}), the left hand sides of
(\ref{Hsub0epsilonode})--(\ref{Hsub4epsilonode}) can now be written
$DH_{n\epsilon}/D\epsilon$.  In particular, (\ref{Hsub0epsilonode}) is
simply $DH_{0\epsilon}/D\epsilon =0$, which immediately gives
$H_{0\epsilon} = C \circ Z_\epsilon$, where $C$ is a function
independent of $\epsilon$.  Substituting $\epsilon=1$ and the boundary
condition shown in Table~\ref{table}, we find $H=C\circ Z$.  Then
substituting $\epsilon=0$, we find $C=K_0$.  In summary,
	\begin{equation}
	H_{0\epsilon} = K_0 \circ Z_\epsilon.
	\label{Hsub0epsilonsolution}
	\end{equation}
In particular, substituting $\epsilon=1$ we obtain 
	\begin{equation}
	H = K_0 \circ Z, \qquad
	K_0 = H \circ Z^{-1}.
	\label{K0solution}
	\end{equation}
This completes the solution of $H_\epsilon$ and $K$ to lowest order.

The second order equation (\ref{Hsub2epsilonode}) can now be written
	\begin{equation}
	\frac{DH_{2\epsilon}}{D\epsilon} =
	\frac{1}{24} 
	\{G_\epsilon,H_{0\epsilon}\}_3. 
	\end{equation}
We use (\ref{DDepsilondef}) in this, compose both sides with
$Z^{-1}_\epsilon$, integrate between $\epsilon$ and $1$, and use the
boundary condition $H_{2\epsilon}=0$ at $\epsilon=1$.  The result is
	\begin{equation}
	H_{2\epsilon} = -\frac{1}{24}
	\int_\epsilon^1 d\epsilon' \,
	\{G_{\epsilon'},H_{0\epsilon'}\}_3 
	\circ Z^{-1}_{\epsilon'} \circ Z_\epsilon.
	\label{Hsub2epsilonsolution}
	\end{equation}
Finally, setting $\epsilon=0$, we have
	\begin{equation}
	K_2 = -\frac{1}{24}
	\int_0^1 d\epsilon \,
	\{G_\epsilon,H_{0\epsilon}\}_3
	\circ Z^{-1}_\epsilon.
	\label{K2solution}
	\end{equation}
Similarly, we solve the fourth order equation (\ref{Hsub4epsilonode}),
finding
	\begin{equation}
	K_4 = -\frac{1}{24} \int_0^1 d\epsilon \,
	\{G_\epsilon,H_{2\epsilon}\}_3 \circ Z^{-1}_\epsilon
	+\frac{1}{1920} \int_0^1 d\epsilon \,
	\{G_\epsilon,H_{0\epsilon}\}_5 \circ Z^{-1}_\epsilon.
	\label{K4solution}
	\end{equation}
Clearly the solutions for $H_{n\epsilon}$ and $K_n$ at any order $n$
can be written in terms of integrals over lower order solutions.

Let us now choose $\hat U$ so that $K$ will be a function of $I$ at
lowest order in $\hbar$.  We shall work backwards, first finding a
canonical transformation $Z$ such that $K_0 = H \circ Z^{-1}$ is a
function of $I$.  We then imbed this in a one parameter family
$Z_\epsilon$, from which we compute $G_\epsilon$, ${\hat G}_\epsilon$,
${\hat U}_\epsilon$, and finally ${\hat U}$.

\subsection{Construction of $Z$ via action-angle variables}

The desired canonical transformation $Z$ can be specified in terms of the
action-angle variables for the original Hamiltonian $H$ and the
harmonic oscillator.  We let $(A,\phi)$ be the action-angle variables
of $H$, according to the standard construction in classical mechanics,
although we note that $H$ may depend on $\hbar$.  The action is
defined as a function of the energy by 
	\begin{equation}
	A(E) = \frac{1}{2\pi}
	\int_{H<E} dp\,dx.
	\label{actiondef}
	\end{equation}
The integral is taken over the interior of the closed curve $H=E$ (a
level set of $H$).  The action vanishes at the fixed point, and is an
increasing function of energy as we move away from it.
Equation~(\ref{actiondef}) is the standard way to write the definition
of the action, but, keeping in mind the warnings of
\ref{functionnotation}, if we wish to think of $A$ as a function on
phase space, that is, a mapping $:{\mathbb R}^2 \to {\mathbb R}$ then
it makes no sense to write $A(E)$.  When we say that the action and
energy are functions of one another, what we mean is
	\begin{equation}
	H=f_0 \circ A,
	\label{f0def}
	\end{equation}
where $f_0:{\mathbb R} \to {\mathbb R}$ is the function of a single
variable expressing the relationship between energy and action.  (The
0 subscript will be explained below.)  The function $f_0$ is
invertible in the region of interest, so $A = f_0^{-1} \circ H$.  Then
(\ref{actiondef}) can be written more properly by picking a point $z$
in the region in question, writing $E=H(z)$, and then writing
	\begin{equation}
	A(z) = (f_0^{-1} \circ H)(z) = f^{-1}_0(E),
	\end{equation}
instead of the left hand side of (\ref{actiondef}).  Having defined
the action $A$, we then define the conjugate angle $\phi$ by standard
means in classical mechanics (through a generating function).  This
involves a choice of origin (a point where $\phi=0$) on each of the
closed curves $H=E$ in the region of phase space under consideration.
This choice is smooth but arbitrary.  We henceforth regard $A$ and
$\phi$ as specific functions $:{\mathbb R}^2 \to {\mathbb R}$. 

Next we introduce the harmonic oscillator action-angle variables
$(I,\theta)$, where $I$ is given by (\ref{Idef}) and $\theta$ is the
conjugate angle (the geometrical polar angle in the phase plane,
increasing in the clockwise direction).  These are regarded as
functions $:{\mathbb R}^2 \to {\mathbb R}$.  Then the canonical
transformation $Z:{\mathbb R}^2 \to {\mathbb R}^2$ is defined by
	\begin{equation}
	\eqalign{
	A &= I \circ Z, \\
	\phi &=\theta \circ Z,}
	\label{Zdef}
	\end{equation}
which uniquely defines $Z$, since $(A,\phi)$, $z=(x,p)$, and
$(I,\theta)$ are invertible functions of each other (in the regions of
interest).  Then, using the definition (\ref{K0solution}) of $K_0$, we
have
	\begin{equation}
	K_0 = H \circ Z^{-1} = f_0 \circ A \circ Z^{-1} = f_0 \circ I,
	\label{K0ofI}
	\end{equation}
that is, $K_0$ is the same function of the harmonic oscillator action
$I$ as $H$ is of its own action $A$.  This is reasonable, since
canonical transformations preserve area, and the action is
proportional to the area inside a level set.  In particular, with this
choice of $Z$, the level sets of $K_0$ are circles.

The smoothness of $Z$ is relevant for the use of the Moyal star
product series (\ref{AstarBseries}) which involves derivatives of
functions on phase space.  We believe that $Z$ is smooth at all points
of the relevant domain, under the assumption that $H$ is smooth and
has a generic fixed point.  If the fixed point is not generic (for
example, the quartic oscillator), then $Z$ need not be smooth.  These
questions are discussed further in \ref{Zissmooth}.

\subsection{Finding $Z_\epsilon$, $G_\epsilon$ and ${\hat U}_\epsilon$}  

Now that we have $Z$, we imbed it in a smooth family $Z_\epsilon$ with
the boundary values shown in Table~\ref{table}.  Sj\"ostrand and
Zworski (2002) show that this can be done in a neighborhood of the
fixed point, and Evans and Zworski (2004) give another proof that
applies in the full domain.  For later reference, we also define
$\epsilon$-dependent versions of the action-angle variables,
	\numparts
	\begin{eqnarray}
	A_\epsilon &= I \circ Z_\epsilon, \label{Aepsilondef}\\
	\phi_\epsilon &=\theta \circ Z_\epsilon,
	\label{phiepsilondef}
	\end{eqnarray}
	\endnumparts
with boundary values shown in Table~\ref{table}.  Then we have
	\begin{equation}
	H_{0\epsilon} = f_0 \circ A_\epsilon.
	\label{H0epsaa}
	\end{equation}
All three Hamiltonians, $H$, $H_{0\epsilon}$ and $K$ are the same
function ($f_0$) of their own actions ($A$, $A_\epsilon$ and $I$,
respectively).  

Next we wish to find a function $G_\epsilon$ such that
(\ref{Zsubepsilonode}) is satisfied for the given $Z_\epsilon$.  This
can always be done, since that equation can be solved for the
derivatives $G_{\epsilon,\mu}$, the components of a closed 1-form
(hence exact, since the region is contractible).  This is a standard
result in classical mechanics (Arnold, 1989), which is summarized in
component language in \ref{Gexists}.  The function $G_\epsilon$ is
determined to within an $\epsilon$-dependent, additive constant.  In
the following we drop this constant, since its only effect is to
introduce an $\epsilon$-dependent phase into ${\hat U}_\epsilon$,
which has no effect on the transformed Hamiltonian.

Finally, given $G_\epsilon$, we transform it into the operator ${\hat
G}_\epsilon$, and then define ${\hat U}_\epsilon$ as the solution of
	\begin{equation}
	\frac{d{\hat U}_\epsilon}{d\epsilon} =
	-\frac{i}{\hbar} {\hat U}_\epsilon {\hat G}_\epsilon,
	\label{Uhatepsilonode}
	\end{equation}
subject to the initial condition ${\hat U}_\epsilon=1$ at
$\epsilon=0$.  Then we set ${\hat U} = {\hat U}_\epsilon$ at
$\epsilon=1$.  This completes the preparatory transformation (the
construction of $\hat U$ such that $\hat K$ has a symbol that is a
function of $I$ at lowest order).  We do not need to solve
(\ref{Uhatepsilonode}) explicitly, since for the purposes of this
paper we only need to calculate the effect on the symbol of a slowly
varying operator when it is conjugated by $\hat U$.  But it is
important to know that $\hat U$ exists, as we have shown.

The preparatory transformation might have been carried out with
oscillatory integrals coming from the integral representation of the
Moyal star product, rather than in terms of a path $Z_\epsilon$
through the group of canonical transformations.  Indeed, we tried this
approach initially, but found that it led to complicated algebra
beyond lowest order that we were not able to organize to our
satisfaction.  Perhaps with more effort that approach could be cast
into suitable form.

The formalism we have presented is slightly simpler if we assume that
the path through the group of canonical transformations, $Z_\epsilon$,
$0\le \epsilon \le 1$, is a one-parameter subgroup, that is, that
$G_\epsilon$ is independent of $\epsilon$.  This, however, is a
special assumption that we did not want to make.  Moreover, the use of
an arbitrary path allows us to study what happens when we vary the
path, which leads to interesting conclusions (see below).

\section{Second stage transformations}

In the second stage (the second arrow in (\ref{HKMstages})) we
transform $\hat K$ into a new Hamiltonian $\hat M$, such that the
symbol $M$ is formally a function of $I$ to all orders in $\hbar$.  We
do this by Lie algebraic (power series) techniques that are similar to
those used in classical perturbation theory (Dragt and Finn, 1976;
Cary 1981), although here there are higher order Moyal brackets
appearing as well as Poisson brackets.  See also Littlejohn and
Weigert (1993) for an example of a Moyal-based perturbation
calculation applied to an adiabatic problem in quantum mechanics.

\subsection{The higher order transformations}
\label{higherordertransformations}

We apply a sequence of near-identity unitary transformations, each of
which is responsible for making the symbol of the Hamiltonian a
function of $I$ at two successive orders of $\hbar$.  Only even powers
of $\hbar$ occur in this process.  The sequence is defined by
	\begin{eqnarray}
	{\hat M}^{(0)} &= {\hat K}, \nonumber \\
	{\hat M}^{(2)} &= {\hat U}_2 \, {\hat M}^{(0)} \,
	{\hat U}_2^\dagger, \nonumber \\
	{\hat M}^{(4)} &= {\hat U}_4 \, {\hat M}^{(2)} \,
	{\hat U}_4^\dagger,
	\label{Mhatordersdef}
	\end{eqnarray}
etc., where
	\begin{equation}
	{\hat U}_n = \exp(-i \hbar^{n-1} {\hat G}_n),
	\label{Undef}
	\end{equation}
and where ${\hat G}_n$ is the $n$-th order generator, assumed to have
a symbol $G_n$ that is slowly varying.  Then, for example, the
expression for ${\hat M}^{(2)}$ can be written as a series in $\hbar$
involving iterated commutators,
	\begin{equation}
	{\hat M}^{(2)} = {\hat K} -i\hbar [{\hat G}_2, 
	{\hat K}] - \frac{\hbar^2}{2}
	[{\hat G}_2,[{\hat G}_2,{\hat K}]] + \ldots,
	\label{M2hatseries}
	\end{equation}
and similarly for ${\hat M}^{(4)}$ etc.  Transcribing
(\ref{M2hatseries}) to symbols and using (\ref{Moyalcommutator}), we
have
	\begin{equation}
	M^{(2)} = K + \hbar^2 \{G_2,K\} +\hbar^4 \left(
	-\frac{1}{24} \{G_2,K\}_3 + \frac{1}{2}
	\{G_2,\{G_2,K\}\}\right) + \ldots.
	\label{M2series}
	\end{equation}
In a similar manner we write out commutator  expansions for the
higher order transformations in (\ref{Mhatordersdef}), transcribe
them into symbols, compose the transformations together, and
substitute the expansion (\ref{Ksubndef}).  We write the result in the
form,
	\begin{equation}
	M=M_0 + \hbar^2 M_2 + \hbar^4 M_4 + \ldots,
	\label{Mexpansion}
	\end{equation}
where $M=M^{(\infty)}$, the symbol of the final Hamiltonian after all
the second stage unitary transformations have been carried out, and
where	
	\numparts
	\begin{eqnarray}
	M_0 &= K_0,  \label{M0eqn} \\
	M_2 &= K_2 + \{G_2,K_0\}, \label{M2eqn} \\
	M_4 &= K_4 + \{G_2,K_2\} 
	-\frac{1}{24} \{G_2,K_0\}_3 \nonumber \\
	&\qquad + \frac{1}{2}
	\{G_2,\{G_2,K_0\}\} + \{G_4,K_0\},
	\label{M4eqn}
	\end{eqnarray}
	\endnumparts
etc.  Each $M_n$ is slowly varying.

We want $M$ to be a function only of $I$.  At lowest order we have
this already,
	\begin{equation}
	M_0 = K_0 = f_0 \circ I = H \circ Z^{-1}.
	\label{M0solution}
	\end{equation}

At second order, we wish to choose $G_2$ in (\ref{M2eqn}) so that
$M_2$ will be a function only of $I$, that is, independent of
$\theta$.  In the next few steps it is convenient to bring back the
abuse of notation rejected in \ref{functionnotation}, and to think of
functions like $K_2$, $M_2$, etc. as functions of either $z=(x,p)$ or
of the action-angle coordinates $(\theta,I)$, as convenient.  Then the
Poisson bracket in (\ref{M2eqn}) can be computed in action-angle
variables, whereupon we have
	\begin{equation}
	M_2 = K_2 + \frac{\partial G_2}{\partial \theta} \,
	\omega(I), 
	\label{M2AAeqn}
	\end{equation}
where $\omega(I) = dK_0/dI$.  Note that as a function, $\omega =
f'_0$, since $K_0 = f_0 \circ I$, so $\omega(A)= dH/dA = f'_0(A)$.
Thus, $\omega(A)$ is the frequency of the classical oscillator with
Hamiltonian $H$.  If we now average both sides of (\ref{M2AAeqn}) over
the angle $\theta$, we obtain
	\begin{equation}
	M_2 = {\bar K}_2,
	\label{M2K2avg}
	\end{equation}
where the overbar represents the $\theta$ average.  The simple result
is that $M_2$ is just the average of $K_2$, given by
(\ref{K2solution}).

Then subtracting (\ref{M2K2avg}) from (\ref{M2AAeqn}) and rearranging,
we obtain
	\begin{equation}
	\frac{\partial G_2}{\partial \theta} =
	-\frac{1}{\omega(I)} {\tilde K}_2,
	\label{G2eqn}
	\end{equation}
where the tilde represents the oscillatory part in $\theta$ of a
function.  Equation~(\ref{G2eqn}) always has a solution $G_2$ that is
a periodic function of $\theta$, that is, it is a single-valued
function of $(x,p)$, since ${\tilde K}_2$ has a Fourier series in
$\theta$ without the constant term.  Thus we have shown that it is
possible to choose $G_2$ in (\ref{M2eqn}) such that $M_2$ is
independent of $\theta$.

The same structure persists at all higher orders.  For example, taking
the averaged and oscillatory parts of the fourth order equation
(\ref{M4eqn}) yields an expression for $M_4$ that is independent of
$\theta$ and a solvable equation for $G_4$.  This shows that it is
possible to transform the original Hamiltonian $\hat H$ into a
function of the harmonic oscillator $\hat I$ to all orders in $\hbar$,
at least in the sense of a formal power series for the symbol.

\subsection{Doing the $\epsilon$-integral}

The following steps require some notation and an important theorem
regarding averaging operators that are explained in
\ref{notationaveraging}.  The theorem in question is (\ref{Favgxfm}),
which we apply to (\ref{M2K2avg}), using (\ref{K2solution}), to obtain
a useful form of the expression for $M_2$:
	\begin{eqnarray}
	M_2 & = -\frac{1}{24} \int_0^1 d\epsilon \,
	\left< \xymatrix@C=1pc{(G_\epsilon \ar@<1.0ex>[r]
	\ar[r] \ar@<-1.0ex>[r] & H_{0\epsilon})} \circ
	Z^{-1}_\epsilon \right>_\theta
	\nonumber \\
	&= -\frac{1}{24} \int_0^1 d\epsilon \,
	\left< \xymatrix@C=1pc{G_\epsilon \ar@<1.0ex>[r]
	\ar[r] \ar@<-1.0ex>[r] & H_{0\epsilon}} 
	\right>_{\phi_\epsilon} \circ Z^{-1}_\epsilon,
	\label{M2rearrangeavg}
	\end{eqnarray}
where the diagrammatic notation is explained in \ref{Moyalnotation}.
The $\epsilon$-integration in (\ref{M2rearrangeavg}) can be done,
yielding an expression independent of $\epsilon$, that is, independent
of the path taken through the group of unitary or canonical
transformations used in the preparatory transformation. 

First we transform the integrand of (\ref{M2rearrangeavg}) as
described in \ref{xfmintegrand}, to obtain
	\begin{eqnarray}
	M_2 &= \frac{1}{24} \int_0^1 d\epsilon \,
	\left[ \frac{d}{dA_\epsilon}\left(
	\frac{1}{\omega\circ A_\epsilon}
	\left< \xymatrix@C=1pc{H_{0\epsilon}
	\ar[r] & G_\epsilon \ar@<0.5ex>[r] 
	\ar@<-0.5ex> [r] & H_{0\epsilon}}
	\right>_{\phi_\epsilon} \right)\right]
	\circ Z^{-1}_\epsilon
	\nonumber \\
	&= \frac{1}{24} \frac{d}{dI} \left( 
	\frac{1}{\omega \circ I}
	\int_0^1 d\epsilon \,
	\left< \xymatrix@C=1pc{H_{0\epsilon}
	\ar[r] & G_\epsilon \ar@<0.5ex>[r] 
	\ar@<-0.5ex> [r] & H_{0\epsilon}}
	\right>_{\phi_\epsilon} 
	\circ Z^{-1}_\epsilon\right)
	\nonumber \\
	&= \frac{1}{24} \frac{d}{dI} \left(
	\frac{1}{\omega \circ I}
	\left< \int_0^1 d\epsilon \,
	\xymatrix@C=1pc{(H_{0\epsilon}
	\ar[r] & G_\epsilon \ar@<0.5ex>[r] 
	\ar@<-0.5ex> [r] & H_{0\epsilon})}
	\circ Z^{-1}_\epsilon
	\right>_\theta\right),
	\label{M2firstxfm}
	\end{eqnarray}
where once we have transformed $A_\epsilon$ into $I$ by composing with
$Z^{-1}_\epsilon$ we can pull the factors depending on it out of the
integral, since they are no longer $\epsilon$-dependent.  Next we use
the methods described in \ref{antideriv} to guess and prove that
	\begin{equation}
	\frac{1}{2}
	\frac{D}{D\epsilon}
	\xymatrix@C=1pc{(H_{0\epsilon} \ar@<0.5ex>[r] \ar@<-0.5ex>[r]
	& H_{0\epsilon})} = 
	\xymatrix@C=1pc{(H_{0\epsilon}
	\ar[r] & G_\epsilon \ar@<0.5ex>[r] 
	\ar@<-0.5ex> [r] & H_{0\epsilon})}.
	\label{HHantideriv}
	\end{equation}
This makes the integral (\ref{M2firstxfm}) easy to do, yielding,
	\begin{equation}
	M_2 =\frac{1}{48} \frac{d}{dI} \left(
	\frac{1}{\omega \circ I}	
	\left< \xymatrix@C=1pc{(H \ar@<0.5ex>[r] \ar@<-0.5ex>[r]
	& H)} \circ Z^{-1} 
	- \xymatrix@C=1pc{(K_0 \ar@<0.5ex>[r] \ar@<-0.5ex>[r]
	& K_0)} \right>_\theta\right).
	\label{M2solution}
	\end{equation}

Let us call the two terms on the right hand side of (\ref{M2solution})
the ``$H$-term'' and the ``$K_0$-term.''  Since $K_0=f_0 \circ I$, the
Moyal bracket in the $K_0$-term can be expanded out by the chain rule
in terms of derivatives of $f_0$ and diagrams involving $I$.  We find
	\begin{equation}
	\xymatrix@C=1pc{(K_0 \ar@<0.5ex>[r] \ar@<-0.5ex>[r]
	& K_0)} = 2f'_0 f''_0 
	\xymatrix@C=1pc{(I \ar[r] & I & I \ar[l])}
	+f^{\prime\,2}_0
	\xymatrix@C=1pc{(I \ar@<0.5ex>[r] \ar@<-0.5ex>[r]
	& I)},
	\label{K0K0expansion}
	\end{equation}
where $f'_0$ means $f'_0 \circ I$, etc., and where some diagrams have
vanished since $\xymatrix@C=1pc{(I \ar[r] & I)} = \{I,I\} = 0$.  The
nonvanishing diagrams can be calculated using (\ref{Idef}), which
gives 
	\begin{equation}
	\xymatrix@C=1pc{(I \ar[r] & I & I \ar[l])}=2I,
	\qquad
	\xymatrix@C=1pc{(I \ar@<0.5ex>[r] \ar@<-0.5ex>[r]
	& I)} = 2,
	\label{Idiagrams}
	\end{equation}
so the $K_0$-term is a function only of $I$ and the angle average in
(\ref{M2solution}) does nothing to this term.  Finally we take the
$I$-derivative and compute the $K_0$-term explicitly, finding,
	\begin{equation}
	\hbox{\rm $K_0$-term} =
	-\frac{f''_0}{8}
	-\frac{f^{\prime\prime\prime}_0}{12}I.
	\label{K0term}
	\end{equation}

The intermediate Hamiltonian $K$ is not unique, because of the choice
of the path $Z_\epsilon$ through the group of canonical
transformations that connects the identity at $\epsilon=0$ and the
given transformation $Z$ at $\epsilon=1$.  More precisely, $K_0 =
H\circ Z^{-1}$ is unique because it is expressed purely in terms of
$Z$, but $K_2$ and all higher order terms depend on $Z_\epsilon$ at
intermediate values of $\epsilon$.  Nevertheless, by (\ref{M2K2avg}),
if we vary the path $Z_\epsilon$ while keeping the endpoints fixed,
$K_2$ can change by at most a function whose $\theta$-average is zero,
so that $M_2$ remains invariant.  Such a function can be written as
the $\theta$-derivative of some other function.  These facts are
proven in \ref{uniquenessofK2}.

\section{The eigenvalues}
\label{eigenvalues}

We have shown how to transform the original Hamiltonian $\hat H$ into
a new Hamiltonian $\hat M$ whose symbol $M$ is a function of $I$ to
any desired order in $\hbar$, and we have explicitly evaluated the
first two terms $M_0$ and $M_2$ of the series for $M$.  Let us write
$M_n = g_n \circ I$, thereby defining the functions $g_n$, so that $M
= g \circ I$, where $g=g_0 + \hbar^2 g_2 + \hbar^4 g_4 + \ldots$.  In
view of (\ref{M0solution}) we have $g_0=f_0$, and $g_2$ is given
implicitly by (\ref{M2solution}).  

As mentioned above, an operator is a function of $\hat I$ if and only
if its symbol is a function of $I$.  The two functions are the same at
lowest order in $\hbar$, but it turns out that they differ at higher
order.  These facts are proved in \ref{symbolofafunction}.  Thus, if
we define a function $f$ by ${\hat M} = f({\hat I})$ and expand it
according to $f=f_0 + \hbar^2 f_2 +\hbar^4 f_4 + \ldots$, then we will
have $f_0=g_0$ but $f_2 \ne g_2$.  Thus $f_0$ defined this way is the
same function introduced above in (\ref{f0def}), and we have ${\hat M}
= f_0({\hat I})$ at lowest order.  This is just what we guessed in the
introduction, and it implies the usual Bohr-Sommerfeld formula, since
the eigenvalues of $\hat I$ are $(n+1/2)\hbar$.

\subsection{The Bohr-Sommerfeld rule to higher order}
\label{BSrule}

To carry the Bohr-Sommerfeld rule to higher order, it is necessary to
find the relation between the symbol of an operator and the symbol of
a function of that operator.  This topic is discussed in
\ref{symbolofafunction}.  In the following we are interested in the
case ${\hat M} = f({\hat I})$ and $M=g \circ I$, so we will identify
$\hat M$ and $\hat I$ with operators $\hat B$ and $\hat A$ of
\ref{symbolofafunction}, respectively.  Then (\ref{Bsymbolexpansion})
gives the relation between functions $f$ and $g$.  Expanding $f$ and
$g$ in even $\hbar$ series as above and using $M_n = g_n \circ I$, we
can write (\ref{Bsymbolexpansion}) in the form,
	\begin{eqnarray}
	&M_0 + \hbar^2 M_2 + \ldots \nonumber \\ 
	&=f_0 + \hbar^2
	\left[ f_2  -\frac{f''_0}{16} 
	\xymatrix@C=1pc{(I \ar@<0.5ex>[r] \ar@<-0.5ex>[r]
	& I)}
	-\frac{f^{\prime\prime\prime}_0}{24}
	\xymatrix@C=1pc{(I \ar[r] & I & I) \ar[l]} \right]
	+\ldots  \nonumber \\
	&= f_0 + \hbar^2 \left( f_2 -\frac{f''_0}{8}
	-\frac{f_0^{\prime\prime\prime}}{12}\right)
	+\ldots,
	\end{eqnarray}
where $f_0$ means $f_0 \circ I$, etc., and where we use
(\ref{Idiagrams}).  This implies $M_0=f_0 \circ I$, which we knew
already, and allows us to solve for $f_2$ by equating the final
quantity in the parentheses with $M_2$ in (\ref{M2solution}).  We see
that the second order correction terms coming from
(\ref{Bsymbolexpansion}) exactly cancel the $K_0$-term (\ref{K0term}),
so that $f_2 \circ I$ is just the $H$-term of (\ref{M2solution}),
	\begin{equation}
	f_2 \circ I = \frac{1}{48} \frac{d}{dI} 
	\left( \frac{1}{\omega \circ I}
	\left< \xymatrix@C=1pc{H \ar@<0.5ex>[r] \ar@<-0.5ex>[r]
	& H} \right>_\phi \circ Z^{-1}\right),
	\label{f2solution}
	\end{equation}
where we use (\ref{altFavgxfm}). 

The eigenvalues of $\hat H$ are the same as the eigenvalues of $\hat
M$, which are given by $f \circ I$ evaluated at $I=(n+1/2)\hbar$, or
as we shall prefer to write it, $f \circ A$ evaluated at
$A=(n+1/2)\hbar$ (in this final step we are starting to confuse the
functions $I$, $A$, with the values $I$, $A$).  We compose $f \circ I
= f_0 \circ I + \hbar^2 f_2 \circ I + \ldots$ with $Z$ and use
(\ref{Zdef}), (\ref{M0solution}) and (\ref{f2solution}) to
obtain (\ref{BS2ndorder}), which is the Bohr-Sommerfeld formula
including $O(\hbar^2)$ corrections.  

Equation~(\ref{BS2ndorder}) is manifestly invariant under linear
canonical transformations, since the matrix $J^{\mu\nu}$ is invariant
under conjugation by a symplectic matrix.  Therefore, although this
equation was derived in coordinates $(x,p)$ with balanced units of
$\hbox{\rm action}^{1/2}$, the original units may be restored by a
canonical scaling transformation, and the answer remains the same.

In the case $H=p^2/2m + V(x)$, we have $\{H,H\}_2 = 2 V''(x)/m$, and
(\ref{BS2ndorder}) agrees with the second order results of Bender and
Orszag (1978), although we omit the details of the comparison.
Equation~(\ref{BS2ndorder}) also agrees with the recent result of
Colin de Verdi\`ere (2004).  We have also derived (\ref{BS2ndorder})
by a completely different method (a kind of WKB-Maslov method), and
obtained the same answer.  We believe (\ref{BS2ndorder}) is correct.

\subsection{Action operators}

The formalism presented naturally suggests a definition of an ``action
operator.''  Let ${\hat V}$ be the overall unitary transformation
resulting from the composition of the preparatory and second stage
transformations,
	\begin{equation}
	{\hat V} = \ldots {\hat U}_4 {\hat U}_2 {\hat U},
	\label{Vhatdef}
	\end{equation}
so that 
	\begin{equation}
	{\hat M} = {\hat V} {\hat H} {\hat V}^\dagger
	= f({\hat I}). 
	\label{MVHeqn}
	\end{equation}
We then define an action operator $\hat B$ by
	\begin{equation}
	{\hat B} = {\hat V}^\dagger {\hat I} {\hat V},	
	\label{Bhatdef}
	\end{equation}
so that 
	\begin{equation}
	{\hat H} = {\hat V}^\dagger f({\hat I}) {\hat V}
	= f({\hat B}).
	\label{HofBeqn}
	\end{equation}
This is the relation whose expression in terms of symbols is the
Bohr-Sommerfeld formula.  It is straightforward to write out the
symbol $B$ of ${\hat B}$ in a power series in $\hbar$.  Our analysis
of the multidimensional Bohr-Sommerfeld formula involves action
operators in a more intimate way than the one-dimensional case.

One can also transform creation and annihilation operators.  Let
${\hat a}= ({\hat x} + i{\hat p})/(\sqrt{2}\hbar)$, ${\hat a}^\dagger =
({\hat x} - i{\hat p})/(\sqrt{2}\hbar)$, so that ${\hat I} = ({\hat
a}^\dagger {\hat a} + 1/2)\hbar$, and define the unitarily equivalent
operators ${\hat b} = {\hat V}^\dagger {\hat a} {\hat V}$, ${\hat
b}^\dagger = {\hat V}^\dagger {\hat a}^\dagger {\hat V}$.   In this
way many of the algebraic relations involving creation and
annihilation operators for the harmonic oscillator go over to more
general oscillators, for example, ${\hat B} = (b^\dagger b +
1/2)\hbar$.

\section{Conclusions}

We conclude by presenting some comments on the present calculation.  

We could have expanded $H$ in a power series in $\hbar$, as in
(\ref{slowlyvaryingdef}), and used the boundary conditions
$H_{n\epsilon}=H_n$ at $\epsilon=1$, which would have made all the
symbols of this paper, $H_{n\epsilon}$, $K_n$, $M_n$, etc.,
independent of $\hbar$.  We did not do this because the odd powers of
$\hbar$ in the expansion of $H$ would complicate all subsequent
formulas without otherwise raising any new, essential issues to be
dealt with.  The essence of the procedure we have given is one that
operates only with even powers of $\hbar$.  

In the calculation above there was a ``miraculous'' cancellation of
the $K_0$-term (\ref{K0term}), where in one instance it arose as a
consequence of doing the $\epsilon$-integral for $K_2$, and in the
second as a consequence of working out the symbol of a function of an
operator.  One suspects that this cannot be accidental.  We will
provide a deeper insight into this cancellation in our subsequent work
on the multidimensional problem.

The derivation of the multidimensional generalization of the
Bohr-Sommerfeld formula (including order $\hbar^2$ corrections), also
known as the Einstein-Brillouin-Keller or torus quantization rule,
requires new diagrammatic methods not considered in this
paper.  The answer is not an obvious generalization of the
one-dimensional formula, and it involves some new geometrical issues
for its interpretation.   These topics will be the subject of a
companion paper.

\ack

The authors would like to acknowledge stimulating and useful
conversations with Yves Colin de Verdi\`ere, Jonathan Robbins, Alan
Weinstein and Maciej Zworski.  We would also like to thank the
Mathematical Sciences Research Institute for their hospitality while
part of this work was being carried out.  Research at MSRI is
supported in part by NSF grant DMS-9810361.
  
\appendix

\section{Notation for Moyal star product}
\setcounter{section}{1}
\label{Moyalnotation}

The Moyal star product $A * B$ of two symbols $A$, $B$ is the symbol
of the operator product ${\hat A}{\hat B}$.  We write the $\hbar$
expansion of this product in the following notation,
	\begin{equation}
	A * B = \sum_{n=0}^\infty 
	\frac{1}{n!}
	\left( \frac{i\hbar}{2}\right)^n
	\{A,B\}_n.
	\label{AstarBseries}
	\end{equation}
We call the bracket $\{\;,\;\}_n$ that occurs in this series the
``$n$-th order Moyal bracket'' (other authors use this terminology to
mean something else).  This bracket is defined as follows.  First, we
define the Poisson tensor and its inverse by means of component
matrices in the $z^\mu=(x,p)$ coordinates,
	\begin{equation}
	J^{\mu\nu} = \left(
	\begin{array}{cc}
	0 & -1 \\
	1 & 0 
	\end{array}
	\right),
	\qquad
	J_{\mu\nu} = \left(
	\begin{array}{cc}
	0 & 1 \\
	-1 & 0
	\end{array}
	\right).
	\label{Jmunudef}
	\end{equation}
Note that $J_{\mu\nu}$ are the components of the symplectic form.  We
use $J^{\mu\nu}$ or $J_{\mu\nu}$ to raise and lower indices.  This
proceeds much as in metrical geometry, but one should note the sign
change in
	\begin{equation}
	X^\mu Y_\mu = - X_\mu Y^\mu.
	\label{contractionsign}
	\end{equation}
(In this paper we sum over repeated indices.)  Next, we define
	\appendixnumparts
	\begin{eqnarray}
	\{A,B\}_0 & = AB, 
	\label{Moyalbracket0def} \\
	\{A,B\}_1 & = A_{,\,\mu} \, J^{\mu\alpha} \, B_{,\,\alpha}, \\
	\{A,B\}_2 & = A_{,\,\mu\nu} \, J^{\mu\alpha} J^{\nu\beta} \,
	  B_{,\,\alpha\beta}, 
	\label{Moyalbracket2def} \\
	\{A,B\}_3 & = A_{,\,\mu\nu\sigma}\, J^{\mu\alpha} 
	J^{\nu\beta} J^{\sigma\gamma} \, B_{,\,\alpha\beta\gamma},
	\label{Moyalbracket3def}
	\end{eqnarray}
	\endappendixnumparts
etc., as required for (\ref{AstarBseries}) to be the standard Moyal
series for the star product.  Note that $\{\;,\;\}_1$ is the usual
Poisson bracket.  In this paper a bracket $\{\;,\;\}$ without a
subscript will be assumed to be a Poisson bracket.  Note also that
	\begin{equation}
	\{A,B\}_n = (-1)^n \{B,A\}_n. 
	\label{ABswapsign}
	\end{equation}
Finally, note that if ${\hat C} = [{\hat A},{\hat B}]$, then the Moyal
series for the symbol of the commutator is
	\begin{eqnarray}
	C & = [A,B]_* = 2 \sum_{n=1,3,5,\ldots}
	\frac{1}{n!}
	\left(\frac{i\hbar}{2}\right)^n
	\{A,B\}_n \nonumber \\
	&= i\hbar\left(
	\{A,B\} - \frac{\hbar^2}{24} \{A,B\}_3
	+ \frac{\hbar^4}{1920} \{A,B\}_5 - \ldots\right),
	\label{Moyalcommutator}
	\end{eqnarray}
which defines the notation $[A,B]_*$.

In this paper we make use of an alternative, diagrammatic
notation for $n$-th order Moyal brackets and related expressions.  For
example, the ordinary Poisson bracket is be written 
	\begin{equation}
	\{A,B\} = \xymatrix@C=1pc{A \ar[r] & B},
	\label{AarrowB}
	\end{equation}
where the arrow indicates differentiations applied to the operands $A$
and $B$, connected by the $J^{\mu\nu}$ tensor.  The base of the arrow
is attached to the first index of $J^{\mu\nu}$ and the tip to the
second index.  The operands can be placed in any position, as long as
the arrow goes in the right direction:
	\begin{equation}
	\xymatrix@C=1pc{A \ar[r] & B}
	=\vcenter{\xymatrix@R=1pc{A \ar[d] \\ B}}
	=\vcenter{\xymatrix@R=1pc{B \\ A \ar[u]}}
	=\xymatrix@C=1pc{B & A \ar[l]}.
	\label{anydirection}
	\end{equation}
But if the direction of the arrow is reversed, then there is a sign
change, due to the antisymmetry of $J^{\mu\nu}$:
	\begin{equation}
	\xymatrix@C=1pc{A \ar[r] &B} = -\xymatrix@C=1pc{(A & B) \ar[l]},
	\end{equation}
which is the usual antisymmetry of the Poisson bracket.  Similarly,
the second Moyal bracket is given by
	\begin{equation}
	\{A,B\}_2 = \xymatrix@C=1pc{A \ar@<0.5ex>[r] \ar@<-0.5ex>[r] & B}
	          =\xymatrix@C=1pc{A & B \ar@<0.5ex>[l] \ar@<-0.5ex>[l]}.
	\end{equation}
The two expressions on the right are equal because changing the
direction of both arrows changes the sign twice.  In this notation,
the Jacobi identity is
	\begin{equation}
	[\xymatrix@C=1pc{(A \ar[r] &B) \ar[r] & C}]+
	[\xymatrix@C=1pc{(B \ar[r] &C) \ar[r] & A}]+
	[\xymatrix@C=1pc{(C \ar[r] &A) \ar[r] & B}]=0,
	\end{equation}
where the square brackets are only for clarity.  The first term can be
expanded out by the chain rule, which in diagrammatic notation gives
	\begin{equation}
	\xymatrix@C=1pc{(A \ar[r] &B) \ar[r] & C}=
	\xymatrix@C=1pc{(A \ar[r] &B \ar[r] &C)} +
	\xymatrix@C=1pc{(C & A\ar[l] \ar[r] & B)}.
	\end{equation}
Similarly expanding the other two terms gives the vanishing sum of six
diagrams, providing a diagrammatic proof of the Jacobi identity.

\section{Notation for functions}
\label{functionnotation}

In this paper it is convenient to use the (slightly nonstandard)
notation $f:A \to B$ to mean that the domain of function $f$ is some
suitably chosen subset of set $A$ (in the standard notation, $A$
itself is the domain).

For the calculations of this paper it is important to avoid the usual
abuse of notation in physics in which a function is confused with the
value of a function.  (Actually it is practically impossible to avoid
this everywhere, but we shall do so wherever it is likely to cause
confusion.)  A ``function'' means a mapping, for example, $H,
H_\epsilon, G_\epsilon, \dots : {\mathbb R}^2 \to {\mathbb R}$, and a
canonical transformation is another mapping, $Z_\epsilon, Z, \ldots :
{\mathbb R}^2 \to {\mathbb R}^2$.  The components $\mu=1,2$ of $Z$ or
$Z_\epsilon$ will be denoted $Z^\mu$ or $Z^\mu_\epsilon$; each of
these is a function $:{\mathbb R}^2 \to {\mathbb R}$.  Functions will
be denoted by bare symbols, $H$, $Z^\mu$, etc., whereas values of
functions will involve the specification of an argument, $H(z)$,
$Z^\mu(z_0)$, etc.  It is also important to distinguish the identity
map ${\rm Id}:{\mathbb R}^2 \to {\mathbb R}^2$ from its value, which
are the coordinates themselves.  The identity map is defined by
	\begin{equation}
	{\rm Id}^\mu(z) = z^\mu.
	\label{idmapdef}
	\end{equation}

One must also be careful about notation for derivatives.  We use comma
notation for derivatives since notation such as $\partial A/\partial
z^\mu$ prejudices the choice of symbol to be used for the argument of
the function.  For example, the notation
	\begin{equation}
	\frac{\partial A}{\partial z^\mu}\bigl(Z(z)\bigr)
	\label{ambiguousderiv}
	\end{equation}
is ambiguous; do we differentiate first and then substitute $Z(z)$ for
the argument, or substitute first and then differentiate?  To avoid
this problem, we write $A,_\mu$ for the derivative of $A$, $A,_\mu
\circ Z$ if we wish to differentiate first and then substitute, and
$(A \circ Z),_\mu$ if we wish to substitute first and then
differentiate, where $\circ$ represents the composition of two
functions.  The latter expression can be expanded by the chain rule,
	\begin{equation}
	(A \circ Z)_{,\,\mu} = (A_{,\,\nu} \circ Z) Z^\nu_{,\,\mu}.
	\label{chainruleexample}
	\end{equation}

Poisson and Moyal brackets defined in
(\ref{Moyalbracket0def})--(\ref{Moyalbracket3def}) always denote
functions.  For example, the notation $\{A(z),B(z)\}$ is meaningless,
because it is only possible to take the Poisson bracket of functions,
not numbers (the values of functions).  On the other hand,
$\{A,B\}(z)$ is meaningful.

\section{The smoothness of $Z$}
\label{Zissmooth}

The Moyal product rule (\ref{AstarBseries}) involves derivatives of
symbols, and is not meaningful as it stands if the symbols are not
smooth.  We are assuming that $H$ is smooth, but there is the question
of the smoothness of the transformation $Z$ as we have constructed it.
We believe that under the our assumptions about $H$, the
transformation $Z$ is smooth, over a domain which is the open interior
of a level set of $H$ surrounding the fixed point but lying inside the
first separatrix.  We have not proved this, but in the following we
present some considerations relevant to the question.  We also present
an example in which some of our assumptions about $H$ are violated and
$Z$ is not smooth.

There are two canonical transformations that are used in the
construction of $Z$, one taking us from $z=(x,p)$ to $(\phi,A)$, and
the other from $z=(x,p)$ to $(\theta,I)$.  The transformation $Z$
defined by (\ref{Zdef}) is the composition of one of these canonical
transformations with the inverse of the other.  These two canonical
transformations are smooth except where $\theta$ or $\phi$ jumps from
0 to $2\pi$, and except at the fixed point, where $I=A=0$ and $\theta$
or $\phi$ is undefined.  Therefore $Z$ is also smooth, except possibly
at these places.  

Let the transformation from $z=(x,p)$ to $(\theta,I)$ be given by
	\begin{equation}
	\eqalign{
	x &= \sqrt{2I} \sin\theta, \\
	p &= \sqrt{2I} \cos\theta,}
	\label{xptothetaI}
	\end{equation}
which amounts to a convention for the origin of the angle $\theta$ (it
lies along the $p$-axis).  This transformation is written without
regard to the warnings of \ref{functionnotation}, but if we think of
$z^\mu = (x,p)$ as values ($\in {\mathbb R}$) and $\theta$ and $I$ as
functions, then the equation is put into proper notation by writing
the left hand side as $\hbox{\rm Id}^\mu(z)$ and $\theta$ and $I$ on
the right hand side as $\theta(z)$ and $I(z)$.  Thus,
(\ref{xptothetaI}) expresses the relation between the functions
$\hbox{\rm Id}^\mu$ and functions $(\theta,I)$.  Now composing this
with $Z$ gives
	\begin{equation}
	Z^\mu = \left(\begin{array}{c}
	\sqrt{2A} \sin \phi \\
	\sqrt{2A} \cos \phi
	\end{array}\right)
	\label{ZmuofAphi}
	\end{equation}
which gives an explicit representation of functions $Z^\mu$ in terms
of functions $A$ and $\phi$.  In particular, this shows that $Z$ is
continuous when $\phi$ jumps from 0 to $2\pi$, as long as $A\ne 0$.  

The inverse transformation can be handled in a similar way.  Let the
transformation from $z^\mu=(x,p)$ to $(\phi,A)$ be expanded in a
Fourier series in $\phi$,
	\begin{equation}
	z^\mu = \sum_n z^\mu_n(A) \, e^{in\phi},
	\end{equation}
where $z^\mu_n: {\mathbb R} \to {\mathbb C}$ are the expansion
coefficients.  This is subject to the same warnings about abuse of
notation as (\ref{xptothetaI}).  When these are straightened out and
the result is composed with $Z^{-1}$, we obtain
	\begin{equation}
	(Z^{-1})^\mu = \sum_n (z^\mu_n \circ I) \, e^{in\theta},
	\end{equation}
an explicit representation of $Z^{-1}$.  

Now we wish to show that $Z$ is smooth at the fixed point.  This means
that $Z$ has derivatives of all orders.  In the following we present
algorithms which we believe correctly give the derivatives of $Z$ and
of other functions at the fixed point, although we do not attempt to
prove this in detail.  

Since $H$ is smooth, it has an expansion about the fixed point,
	\begin{equation}
	H(z)= \frac{1}{2} Q_{\mu\nu} \, z^\mu z^\nu + \ldots,
	\label{Hexpansion}
	\end{equation}
where for simplicity we assume that the fixed point is at $z=0$ and
that the constant term in the expansion vanishes, and where
$Q_{\mu\nu}$ is the positive definite Hessian matrix
(\ref{commanotationdef}) evaluated at the fixed point.  This series
need not converge, but all coefficients in the series (the derivatives
of $H$) are defined.  It is convenient to manipulate such power series
in a formal manner, since the rules for manipulating power series
(multiplying, inverting, composing, etc.) are equivalent to the rules
(chain, Leibnitz, etc.) for expressing the derivatives of new functions in
terms of given derivatives of old ones. 

Birkhoff normal form theory (Birkhoff, 1927; Dragt and Finn, 1976;
Eckhardt, 1986) is a convenient way of developing a power series
expansion of the transformation $Z$.  This theory takes a Hamiltonian
represented as a power series in $z=(x,p)$, whose leading term is a
harmonic oscillator, and transforms it into a function of the harmonic
oscillator action $I$.  The expansion (\ref{Hexpansion}) of $H$ does
not begin with a harmonic oscillator, but can be brought into this
form by means of a linear canonical transformation $L:{\mathbb R}^2
\to {\mathbb R}^2$, as shown by the theory of normal forms for
quadratic Hamiltonians (Arnold, 1989).  That is, there exists linear
symplectic map $L$ such that
	\begin{equation}
	(H \circ L)(z) = \frac{1}{2} a_1 (x^2 + p^2) + \ldots,
	\label{HLexpansion}
	\end{equation}
where $a_1 > 0$, and where the ellipsis represents cubic and higher
terms in a power series.  In this step we rely on the positive
definiteness of $Q_{\mu\nu}$.  Then Birkhoff normal form theory
provides another (nonlinear) canonical transformation $N:{\mathbb R}^2
\to {\mathbb R}^2$, represented as a power series in $z$ in which the
leading (linear) term is the identity transformation, such that $H
\circ L \circ N$ is a function of $I$,
	\begin{equation}
	H \circ L \circ N = f_0 \circ I = a_1 I + a_2 I^2 + \ldots,
	\label{HLNequation}
	\end{equation}
where $f_0$ is the same function introduced in (\ref{actiondef}) and
(\ref{f0def}).  The coefficients $a_1$, $a_2$, etc.\ are determined by
Birkhoff normal form theory, and they give the derivatives of
function $f_0$, which is smooth at the fixed point.  We remark that in
a one-dimensional problem such as this one, there are no resonance
conditions so the Birkhoff algorithm can be carried to any order in
the power series.  

It follows from (\ref{HLNequation}) and (\ref{f0def}) that $A = I
\circ Y$, where $Y=(L \circ N)^{-1}$, so $Y$ has the same effect on
$I$ as does $Z$ in (\ref{Zdef}).  This does not mean that $Y=Z$,
because $Y$ does not necessarily satisfy the second of
equations~(\ref{Zdef}).  But $\theta \circ Y$ is an angle variable
conjugate to $A$, so it differs from $\phi$ only by some phase shift
$\delta$ that depends on $A$.  Assuming this phase shift is well
behaved at $A=0$ (this is really an assumption of reasonableness on
the definition of $\phi$), there exists another canonical
transformation $S$ smooth at $z=0$ so that $Z= Y \circ S$ satisfies
both halves of (\ref{Zdef}).  In this way all the derivatives of $Z$
at $z=0$ may be computed.

We note that if our conditions on $H$ are not met, then $Z$ need not
be smooth at the fixed point.  For example, the relation between
action and energy for the quartic oscillator ($V(x)=x^4$) is given by
$H=cA^{4/3}$, so $K_0 = H \circ Z^{-1} = c' (x^2+p^2)^{4/3}$, where $c$
and $c'$ are constants.  Thus, $K_0$ is not smooth at the fixed point,
and neither is $Z$.

\section{Notation for averaging operators}
\label{notationaveraging}

This appendix develops abuse-free notation for the averaging operator
introduced in subsection~\ref{higherordertransformations}.  Let
$Q:{\mathbb R}^2 \to {\mathbb R}$ be a function on phase space,
treated as a Hamiltonian with evolution parameter $\alpha$,
	\begin{equation}
	\frac{dz^\mu}{d\alpha} = J^{\mu\nu} Q_{,\,\nu}(z),
	\label{QHamseqns}
	\end{equation}
where we assume $Q$ is independent of $\alpha$ so the equations are
autonomous (unlike the case of $G_\epsilon$ considered above).  Let
$Y^Q_\alpha:{\mathbb R}^2 \to {\mathbb R}^2$ be the associated
flow, with components $(Y^Q_\alpha)^\mu$.  The superscript $Q$
indicates the Hamiltonian function generating the flow.  The flow
functions satisfy
	\begin{equation}
	\frac{d(Y^Q_\alpha)^\mu}{d\alpha} =
	J^{\mu\nu} (Q_{,\,\nu} \circ Y^Q_\alpha) =
	\{(Y^Q_\alpha)^\mu, Q\}.
	\label{Yalphaode}
	\end{equation}
We will be interested in the case that $Q$ is an action variable, $I$,
$A_\epsilon$, or $A$.

For example, with $Q=I$, we have an advance map $Y^I_\alpha$ that
advances the angle $\theta$ by $\alpha$.  That is, if a point $z$ of
phase space has action-angle coordinates $(\theta,I)$, then the point
$Y^I_\alpha(z)$ has coordinates $(\theta+\alpha,I)$.  Thus averaging
over the angle $\theta$ can be written as
	\begin{equation}
	{\bar F} = \int_0^{2\pi} 
	\frac{d\alpha}{2\pi} 
	F \circ Y^I_\alpha = \langle F \rangle_\theta,
	\label{Favgthetadef}
	\end{equation}
which defines the notation $\langle F \rangle_\theta$.  Similarly, we
define $\langle F \rangle_{\phi_\epsilon}$ and $\langle F
\rangle_\phi$, using the advance maps $Y^{A_\epsilon}_\alpha$ and
$Y^A_\alpha$. 

The advance maps $Z_\epsilon$ and $Y_\alpha$ are related by the
following identity:
	\begin{equation}
	Z_\epsilon \circ Y^{A_\epsilon}_\alpha 
	= Y^I_\alpha \circ Z_\epsilon.
	\label{ZYidentity}
	\end{equation}
In other words, angle evolution and $\epsilon$-evolution commute.
We prove this by regarding both sides as functions of $\alpha$ at
fixed $\epsilon$, and writing $X_\alpha$ and $X'_\alpha$ for the left
and right hand sides, respectively.  Note that $X_\alpha = X'_\alpha$
at $\alpha=0$.  The left hand side satisfies the differential
equation,
	\begin{eqnarray}
	\frac{dX_\alpha}{d\alpha} &= 
	(Z_{\epsilon,\nu} \circ Y^{A_\epsilon}_\alpha)
	\frac{d(Y^{A_\epsilon}_\alpha)^\nu}{d\alpha} =
	(Z_{\epsilon,\nu} \circ Y^{A_\epsilon}_\alpha)
	\{ (Y^{A_\epsilon}_\alpha)^\nu,A_\epsilon\} 
	\nonumber \\
	&=\{Z_\epsilon \circ Y^{A_\epsilon}_\alpha, A_\epsilon\}
	= \{X_\alpha, A_\epsilon\},
	\label{Xlefteqn}
	\end{eqnarray}
where we have used the chain rule property of the Poisson bracket.
The right hand side satisfies
	\begin{equation}
	\frac{dX'_\alpha}{d\alpha} =
	\frac{dY^I_\alpha}{d\alpha} \circ Z_\epsilon =
	\{Y^I_\alpha,I\} \circ Z_\epsilon =
	\{X'_\alpha,A_\epsilon\},
	\label{Xrighteqn}
	\end{equation}
where we have used (\ref{PBcompositionrule}) and (\ref{Aepsilondef}).
Since $X_\alpha$ and $X'_\alpha$ satisfy the same differential
equation and the same initial conditions, they are equal, $X_\alpha =
X'_\alpha$, and the identity (\ref{ZYidentity}) is proven.  It can
also be written in the form,
	\begin{equation}
	Y^{A_\epsilon}_\alpha \circ Z^{-1}_\epsilon =
	Z^{-1}_\epsilon \circ Y^I_\alpha.
	\label{altZYidentity}
	\end{equation}

It follows from this that for any function $F$ on phase space,
	\begin{equation}
	\left< F \circ Z^{-1}_\epsilon \right>_\theta =
	\left< F \right>_{\phi_\epsilon} \circ Z^{-1}_\epsilon.
	\label{Favgxfm}
	\end{equation}
In view of (\ref{phiepsilondef}) this is a plausible identity.  In
particular, at $\epsilon=1$ we have
	\begin{equation}
	\left<F \circ Z^{-1} \right>_\theta =
	\left<F \right>_\phi \circ Z^{-1}.
	\label{altFavgxfm}
	\end{equation}
To prove (\ref{Favgxfm}), we express the left hand side as an
integral, and then apply (\ref{altZYidentity}):
	\begin{equation}
	\int_0^{2\pi} \frac{d\alpha}{2\pi} \,
	F \circ Z^{-1}_\epsilon \circ Y^I_\alpha =
	\int_0^{2\pi} \frac{d\alpha}{2\pi} \,
	F \circ Y^{A_\epsilon}_\alpha \circ Z^{-1}_\epsilon.
	\end{equation}

\section{Function $G_\epsilon$ exists}
\label{Gexists}

Let $Z_\epsilon^\mu$ be an $\epsilon$-dependent canonical
transformation, defined on a contractible region.  We wish to show
that there exists a function $G_\epsilon$ such that
(\ref{Zsubepsilonode}) is satisfied.  Write $S^\mu{}_\nu =
Z^\mu_{\epsilon,\nu}$ for the derivatives of $Z_\epsilon$, which form
a symplectic matrix.  Then
	\begin{equation}
	\frac{dZ^\mu_\epsilon}{d\epsilon} =
	S^\mu{}_\alpha J^{\alpha\beta} G_{\epsilon,\beta},
	\end{equation}
or
	\begin{equation}
	G_{\epsilon,\beta} = J_{\beta\alpha} (S^{-1})^\alpha{}_\mu
	\frac{dZ^\mu_\epsilon}{d\epsilon}
	= S^\alpha{}_\beta J_{\alpha\mu} 
	\frac{dZ^\mu_\epsilon}{d\epsilon}
	\end{equation}
where we use the property of symplectic matrices, $S^tJS = J$, where
$J$ is the matrix with components $J_{\mu\nu}$.  We must show that the
second derivatives $G_{\epsilon,\beta\gamma}$ are symmetric.
Differentiating, we find
	\begin{equation}
	G_{\epsilon,\beta\gamma} = S^\alpha{}_{\beta,\gamma}
	J_{\alpha\mu}
	\frac{dZ^\mu_\epsilon}{d\epsilon} +
	S^\alpha{}_\beta J_{\alpha\mu}
	\frac{dS^\mu{}_\gamma}{d\epsilon}.
	\end{equation}
The first term on the right hand side is symmetric in
$(\beta,\gamma)$, since
	\begin{equation}
	S^\alpha{}_{\beta,\gamma} =
	Z^\alpha_{\epsilon,\beta\gamma},
	\end{equation}
and the second term is also, as we see by differentiating $S^tJS = J$
with respect to $\epsilon$ and juggling indices.  Thus, the function
$G_\epsilon$ exists.

\section{A transformation of the integrand of (\ref{M2rearrangeavg})}
\label{xfmintegrand}

In this Appendix to save writing we drop the $\epsilon$ subscripts on
$H_{0\epsilon}$, $G_\epsilon$, $A_\epsilon$ and $\phi_\epsilon$,
writing simply $H_0$, $G$, $A$ and $\phi$.  The latter symbols,
however, are not to be confused with the notation indicated in
Table~\ref{table} at $\epsilon=1$.  We have placed this part of the
calculation in an Appendix, to avoid confusion due to the notational
change.

Let us pick out the $\phi$-average of the Moyal bracket in the
integrand of (\ref{M2rearrangeavg}) and write it in an obvious
notation,
	\begin{equation}
	\left< \xymatrix@C=1pc{G \ar@<1.0ex>[r]
	\ar[r] \ar@<-1.0ex>[r] & H_0} 
	\right>_\phi =
	\int_0^{2\pi} \frac{d\phi}{2\pi} \,
	\xymatrix@C=1pc{(G \ar@<1.0ex>[r]
	\ar[r] \ar@<-1.0ex>[r] & H_0)},
	\label{avgMB3}	
	\end{equation}
where the parentheses are only for clarity.  We now introduce a
technique for ``breaking a bond'' of an angle-averaged graph that is
sometimes useful.  The average of course depends only on $A$, if we
think of it as a function of $(\phi,A)$.  We imagine evaluating this
average at constant action $A=a$, which we enforce by inserting a
$\delta$-function and integrating over both $A$ and $\phi$.  This
transforms (\ref{avgMB3}) into
	\begin{equation}
	\int \frac{dA\,d\phi}{2\pi} \,
	\delta(A-a) \,
	\xymatrix@C=1pc{(G \ar@<1.0ex>[r]
	\ar[r] \ar@<-1.0ex>[r] & H_0)},
	\end{equation}
where the integral is taken over a region of phase space that includes
the level set $A=a$ (an orbit of $H_0$).  We then transform variables
of integration to $z=(x,p)$, we use $dA \, d\phi = d^2z$ (since the
transformation is canonical), we write out one of the bonds
explicitly, and we integrate by parts in the variable $z^\nu$:
	\begin{eqnarray}
	&\int \frac{d^2z}{2\pi} \,
	\delta(A-a) \, 
	\xymatrix@C=1pc{(G_{,\,\mu} \ar@<0.5ex>[r]
	\ar@<-0.5ex>[r] & H_{0,\nu})} J^{\mu\nu}
	=-\int \frac{d^2z}{2\pi}  \Bigl[
	\delta'(A-a) \, A_{,\,\nu} 
	\xymatrix@C=1pc{(G_{,\,\mu} \ar@<0.5ex>[r]
	\ar@<-0.5ex>[r] & H_0)} 
	\nonumber \\
	&\qquad +\delta(A-a) \, 
	\xymatrix@C=1pc{(G_{,\,\mu\nu} \ar@<0.5ex>[r]
	\ar@<-0.5ex>[r] & H_0)}\Bigr]
	J^{\mu\nu}.
	\end{eqnarray}
The second term in the final integral vanishes, due to the symmetry of
$G_{,\,\mu\nu}$ and the antisymmetry of $J^{\mu\nu}$.  In the first
term we switch variables of integration back to $(\phi,A)$ and do the
$A$-integration, which gives
	\begin{equation}
	\int \frac{d\phi}{2\pi} \,
	\frac{\partial}{\partial A}
	\xymatrix@C=1pc{(A & G \ar[l] \ar@<0.5ex>[r]
	\ar@<-0.5ex>[r] & H_0)} =
	-\frac{d}{dA} \left(\frac{1}{\omega \circ A}
	\left<
	\xymatrix@C=1pc{H_0 \ar[r] & G \ar@<0.5ex>[r]
	\ar@<-0.5ex>[r] & H_0}\right>_\phi\right),
	\end{equation}
where we use $H_0 = f_0 \circ A$, that is, (\ref{H0epsaa}), and
$\omega = f'_0$, and change the direction of an arrow in the final
form.

\section{An antiderivative for the integral (\ref{M2firstxfm})}
\label{antideriv}

In this Appendix we use the same notational simplifications as in
\ref{xfmintegrand}.  

In guessing an antiderivative that will allow us to do the integral
(\ref{M2firstxfm}), we must express the diagram $\xymatrix@C=1pc{H_0
\ar[r] & G \ar@<0.5ex>[r] \ar@<-0.5ex>[r] & H_0}$, which contains one
$G$ and three arrows, as $D/D\epsilon$ of some other diagram.  We note
by (\ref{altDDepsilondef}) that taking $D/D\epsilon$ of a diagram
introduces both $G$ and an extra arrow.  Therefore taking the
antiderivative must remove $G$ and one arrow.  The only diagram we can
form from two copies of $H_0$ and two arrows is $\xymatrix@C=1pc{H_0
\ar@<0.5ex>[r] \ar@<-0.5ex>[r] & H_0}$, so we compute,
	\begin{equation}
	\frac{1}{2}
	\frac{D}{D\epsilon}
	\xymatrix@C=1pc{(H_0 \ar@<0.5ex>[r] 
	\ar@<-0.5ex>[r] & H_0)} =
	\left(\frac{dH_0}{d\epsilon}
	\xymatrix@C=1pc{\ar@<0.5ex>[r] \ar@<-0.5ex>[r] 
	& H_0}\right) -\frac{1}{2}
	\xymatrix@C=1pc{(H_0 \ar@<0.5ex>[r] 
	\ar@<-0.5ex>[r] & H_0) \ar[r] &G}.
	\label{antideriveq1}
	\end{equation}
The first term can be written,
	\begin{eqnarray}
	\xymatrix@C=1pc{(H_0 \ar[r] & G) \ar@<0.5ex>[r] 
	\ar@<-0.5ex>[r] &H_0} &=
	\xymatrix@C=1pc{(H_0 & H_0 \ar@<0.5ex>[l] 
	\ar@<-0.5ex>[l] \ar[r] & G)}
	+2\left(
	\vcenter{\xymatrix@C=-0.3pc@R=0.8pc{& H_0 \\
	H_0 \ar[rr] \ar[ur] && G \ar[ul]}}\right)
	\nonumber \\
	&+ \xymatrix@C=1pc{(H_0 \ar[r] & G
	\ar@<0.5ex>[r] \ar@<-0.5ex>[r] & H_0)},
	\label{antideriveq2}
	\end{eqnarray}
where we use (\ref{Hsub0epsilonode}) and the chain rule, while in the
second term of (\ref{antideriveq1}) removing the parentheses provides
a factor of $2$, thereby cancelling the first term on the right hand
side of (\ref{antideriveq2}).  As for the triangle diagram, it
vanishes, as we note by writing,
	\begin{equation}
	\left( \vcenter{
	\xymatrix@C=0pc@R=0.8pc{& G \ar[dr] \\
	H_0 \ar[ur] \ar[rr] && H_0}} \right) =
	\left( \vcenter{
	\xymatrix@C=0pc@R=0.8pc{& G \ar[dl] \\
	H_0 && H_0 \ar[ul] \ar[ll]}} \right) =
	-\left( \vcenter{
	\xymatrix@C=0pc@R=0.8pc{& G \ar[dr] \\
	H_0 \ar[ur] \ar[rr] && H_0}} \right),
	\end{equation}
where in the first step we reflect about the vertical line and in the
second reverse the directions of all three arrows.  The overall result
is (\ref{HHantideriv}).

\section{The uniqueness of the intermediate Hamiltonian $K$}
\label{uniquenessofK2}

In this appendix we study how the intermediate Hamiltonian $K$ changes
when the path $Z_\epsilon$ through the space of canonical
transformations is varied.  To do this we compose $Z_\epsilon$ with a
near-identity, $\epsilon$-dependent canonical transformation that
becomes the identity at $\epsilon=0,1$. This is equivalent to
replacing $Z^\mu_\epsilon$ with $Z^\mu_\epsilon + \delta
Z^\mu_\epsilon$, where $\delta Z^\mu_\epsilon = \{Z^\mu_\epsilon,
F_\epsilon\}$, where $F_\epsilon$ is a small, $\epsilon$-dependent
function such that $F_\epsilon=0$ at $\epsilon=0,1$.  The
corresponding variation in the inverse function
$(Z^{-1}_\epsilon)^\mu$ can be found by varying $Z^{-1}_\epsilon \circ
Z_\epsilon = \hbox{\rm Id}$, which gives
	\begin{equation}\
	\delta (Z^{-1}_\epsilon)^\mu = -J^{\mu\nu} \,
	F_{\epsilon,\nu} \circ Z^{-1}_\epsilon.
	\label{deltaZinverse}
	\end{equation}
Then we vary (\ref{Hsub0epsilonsolution}) to obtain,
	\begin{equation}
	\delta H_{0\epsilon} = \{H_{0\epsilon},F_\epsilon \}.
	\label{deltaHsub0epsilon}
	\end{equation}
Finally, to get $\delta G_\epsilon$, we vary (\ref{Zsubepsilonode}) to
obtain,
	\begin{equation}
	\frac{d}{d\epsilon}(\delta Z^\mu_\epsilon) =
	\{\delta Z^\mu_\epsilon, G_\epsilon\} +
	\{Z^\mu_\epsilon, \delta G_\epsilon \}.
	\end{equation}
The first term on the right hand side is $\{\{Z^\mu_\epsilon,
F_\epsilon \}, G_\epsilon\}$, while the left hand side is
	\begin{equation}
	\frac{d}{d\epsilon}
	\{Z^\mu_\epsilon, F_\epsilon\} =
	\{\{ Z^\mu_\epsilon, G_\epsilon\}, F_\epsilon\} +
	\left\{Z^\mu_\epsilon, \frac{dF_\epsilon}{d\epsilon}
	\right\}.
	\end{equation}
Rearranging this and using the Jacobi identity gives
	\begin{equation}
	\left\{ Z^\mu_\epsilon,
	\delta G_\epsilon -\frac{DF_\epsilon}{D\epsilon} 
	\right\}=0,
	\end{equation}
where we use (\ref{altDDepsilondef}), or,
	\begin{equation}
	\delta G_\epsilon = \frac{DF_\epsilon}{D\epsilon},
	\label{deltaGepsilon}
	\end{equation}
where we drop a possible $\epsilon$-dependent constant.

In the next few steps we adopt the same notational simplification
mentioned at the beginning of \ref{xfmintegrand}, and in addition we
drop the $\epsilon$ subscript on $F_\epsilon$ and $Z_\epsilon$.  Then
we combine (\ref{K2solution}), (\ref{deltaZinverse}),
(\ref{deltaHsub0epsilon}) and (\ref{deltaGepsilon}) to obtain,
	\begin{eqnarray}
	\delta K_2 = -\frac{1}{24} \int^1_0 d\epsilon &\Bigg[
	\left\{ \frac{dF}{d\epsilon}, 
	H_0\right\}_3
	-\{\{F, G\}, H_0\}_3+\{G, \{H_0, F\}\}_3
	\nonumber \\
	&\quad -\{\{G, H_0\}_3, F\}
	\Bigg] \circ Z^{-1}.
	\label{deltaK2firstform}
	\end{eqnarray}
In this integral we perform an integration by parts, specified by
	\begin{eqnarray}
	\frac{D}{D\epsilon} \{F, H_0\}_3 &=
	\left\{ \frac{d F}{d\epsilon}, H_0
	\right\}_3 
	+\{F, \{H_0, G\}\}_3
	- \{\{ F, H_0\}_3, G\}
	\nonumber \\
	&= \left[ \frac{d}{d\epsilon} \left(
	\{F, H_0\}_3 \circ Z^{-1}
	\right)\right] \circ Z,
	\end{eqnarray}
which allows us to replace the first term of (\ref{deltaK2firstform})
with an exact $\epsilon$-derivative plus two more terms.  The exact
derivative can be integrated, giving zero because of the boundary
conditions on $F$.  What remains is
	\begin{eqnarray}
	\delta K_2 = -\frac{1}{24} \int^1_0 d\epsilon &\Bigl[
	-\{F, \{H_0,G\}\}_3 +
	\{\{F, H_0\}_3, G\} -\{\{F, G\}, H_0\}_3
	\nonumber \\
	&\quad +\{G, \{H_0, F\}\}_3
	-\{\{G, H_0\}_3, F\}
	\Bigr] \circ Z^{-1}.
	\label{deltaK2secondform}
	\end{eqnarray}

We now use an identity related to the Jacobi identity for operators,
itself a consequence of the associativity of operator multiplication.
Let $\hat A$, $\hat B$ and $\hat C$ be any three operators, and write
out the Jacobi identity $[{\hat A}, [{\hat B}, {\hat C}]] + \hbox{\rm
cyclic}=0$ in symbol form, expanding star commutators according to
(\ref{Moyalcommutator}).  The leading order term is the Jacobi
identity for the Poisson bracket, and the next correction term is
	\begin{equation}
	\{A, \{B, C\}_3\} + \{A, \{B, C\}\}_3
	+\hbox{\rm cyclic} = 0.
	\end{equation}
Using this in (\ref{deltaK2secondform}) allows us to write the
integrand as
	\begin{eqnarray}
	\{H_0, \{G, F\}_3\} \circ Z^{-1} &=
	\left[(\omega \circ A) \{ A, \{G,F\}_3\} \right]
	\circ Z^{-1}
	\nonumber \\
	&= - (\omega \circ I) \frac{\partial}{\partial\theta}
	\left( \{G, F\}_3 \circ Z^{-1}\right),
	\end{eqnarray}
where we have expanded the Poisson bracket with $H_0$ in action-angle
variables.  Finally, on restoring the $\epsilon$'s we have
	\begin{equation}
	\delta K_2 = \frac{1}{24}
	(\omega \circ I) \frac{\partial}{\partial \theta}
	\int^1_0 d\epsilon 
	\{G_\epsilon, F_\epsilon\}_3 \circ Z^{-1}_\epsilon.
	\end{equation}
The variation in $K_2$ is an exact $\theta$-derivative, as claimed,
and $M_2$ is invariant under variations in the path $Z_\epsilon$.

We do not know whether the space of symplectomorphisms we are
considering is simply connected, but if not there arises the
possibility of distinct paths $Z_\epsilon$ that are not homotopic.
Since $M_2$ is unique, it must be that the difference in $K_2$ along
such paths is still an exact $\theta$-derivative.

\section{Functions of operators vs. functions of symbols}
\label{symbolofafunction}

In this appendix we calculate the symbol of a function of an operator,
in terms of the symbol of that operator, as a power series in $\hbar$.
We briefly describe a Green's function approach to this problem, which
as far as we know was first presented by Voros (1977) and which is
discussed further by Colin de Verdi\`ere (2004).  In this appendix
we adopt a general notation, in which $\hat A$ is any Hermitian
operator, $f$ is any function $:{\mathbb R} \to {\mathbb R}$, and
${\hat B} = f({\hat A})$.  The problem will be to find the symbol $B$
in terms of the symbol $A$.

Let $a \in {\mathbb C}$ and let ${\hat G}_a = 1/(a-{\hat A})$ be the
Green's operator associated with $\hat A$.  The symbol $G_a$ of ${\hat
G}_a$ may be computed by demanding $G_a * (a-A) = (a-A) *
G_a = 1$, expanding $G_a = G_{a0} + \hbar G_{a1} + \hbar^2 G_{a2} +
\ldots$, expanding the Moyal star product, and collecting things by
orders in $\hbar$.  One finds that only even powers of $\hbar$ occur
in the expansion of $G_a$, and that otherwise it is easy to solve for
the leading terms.  Through second order, the results are
	\appendixnumparts
	\begin{eqnarray}
	G_{a0} &= \frac{1}{a-A}, 
	\label{G0asolution} \\
	G_{a2} &= -\frac{1}{8} \frac{1}{a-A}
	\left\{ \frac{1}{a-A},A \right\}_2 
	\nonumber \\ & =
	-\frac{1}{8} \left[
	\frac{\xymatrix@C=1pc{(A \ar@<0.5ex>[r] \ar@<-0.5ex>[r]
	& A)}}{(a-A)^3} 
	+ 2 \frac{\xymatrix@C=1pc{(A \ar[r] & A & A) \ar[l]}}
	{(a-A)^4}\right].
	\label{G2asolution}
	\end{eqnarray}
	\endappendixnumparts
This is a special case of the symbol of a function of an operator.
For the general case, write ${\hat B} =f({\hat A})$ in the form,
	\begin{equation}
	{\hat B} = \int_\Gamma \frac{da}{2\pi i}
	\frac{f(a)}{a- {\hat A}},
	\label{Bhatcontourint}
	\end{equation}
where the contour $\Gamma$ runs from $-\infty$ to $+\infty$ just below
the real axis, and then returns just above it.  On taking symbols of
both sides, this becomes
	\begin{equation}
	B = \int_\Gamma \frac{da}{2\pi i} f(a) G_a,
	\label{Bcontourint}
	\end{equation}
or, on substituting the expansion for $G_a$ and doing the integrals,
	\begin{equation}
	B = f(A) -\hbar^2 \left[
	\frac{f''(A)}{16} 
	\xymatrix@C=1pc{(A \ar@<0.5ex>[r] \ar@<-0.5ex>[r]
	& A)} +
	\frac{f^{\prime\prime\prime}(A)}{24}
	\xymatrix@C=1pc{(A \ar[r] & A & A) \ar[l]}\right]
	+ O(\hbar^4).
	\label{Bsymbolexpansion}
	\end{equation}
The Green's function method becomes tedious at higher orders, but
recently Gracia-Saz (2004) has found convenient methods for
calculating the higher order terms, including the multidimensional
case.  It turns out that the fourth order term in
(\ref{Bsymbolexpansion}) contains 13 diagrams.  The Green's function
derivation of (\ref{Bsymbolexpansion}) has required $f$ to be analytic
in a strip around the real axis, but Gracia-Saz has shown that the
same expansion holds more generally.

It was stated above that an operator is a function of $\hat I$ if and
only if the symbol is a function of $I$.  We prove this by noting that
an operator is a function of $\hat I$ if and only if it commutes with
the unitary operator ${\hat U}(t)=\exp(-it{\hat I}/\hbar)$ for all
$t$.  This follows since the spectrum of ${\hat I}$ is nondegenerate.
But the unitary operator ${\hat U}(t)$ is a metaplectic operator
(Littlejohn, 1986), so when we conjugate an operator, ${\hat A} \mapsto
{\hat U}(t) {\hat A} {\hat U}^\dagger(t)$, the symbol $A$ is rotated
in phase space.  Therefore an operator commutes with all ${\hat U}$ if
and only if its symbol is rotationally invariant in phase space, that
is, is a function of $I$.

The same thing can be proven at the level of $\hbar$ expansions.  The
general term of the series (\ref{Bsymbolexpansion}) involves diagrams
composed of copies of $A$ connected by arrows.  But if $A=I$, then all
diagrams with three or more arrows attached to any $I$ vanish, since
$I$ is a quadratic function of $z$.  Therefore the only nonvanishing
diagrams are linear ones and circular ones.  A linear diagram with $n$
$I$'s (two on the ends and $n-2$ in the middle) vanishes if $n$ is
even, and is $2(-1)^{(n-1)/2}I$ if $n$ is odd.  A circular diagram
with $n$ $I$'s vanishes if $n$ is odd, and is $2(-1)^{n/2}$ if $n$ is
even.  Equation~(\ref{Idiagrams}) is a special case of these rules.
For now the point is that both these diagrams are functions of $I$.
Thus the entire series (\ref{Bsymbolexpansion}) is a function of $I$,
for any function $f$.

\section*{References}
\begin{harvard}

\item[] Arnold V I 1989 {\it Mathematical Methods of Classical
Mechanics} (New York: Springer-Verlag) Appendix 6.

\item[] Balazs N L and Jennings B K 1984 {\it Phys. Reports} {\bf 104}
347

\item[] Bayen F, Flato M, Fronsdal C, Lichnerowicz A and Sternheimer
D 1978 {\it Ann. Phys.} {\bf 111} 61

\item[] Bender C M and Orszag S A 1978 {\it Advanced Mathematical Methods for
Scientists and Engineers} (New York: McGraw-Hill)

\item[] Berry M V 1977 {\it Phil. Trans. Roy. Soc.} {\bf 287} 237

\item[] Birkhoff G D 1927 {\it Dynamical Systems} (Providence, Rhode
Island: American Mathematical Society)

\item[] Braam P J and Duistermaat J J 1995 {\it Panoramas of
Mathematics}, Banach Center Publications v.~34 (Warsaw: Institute of
Mathematics, Polish Academy of Sciences) p.~29

\item[] Cary J R 1981 {\it Phys. Rep.} {\bf 79} 131

\item[] Colin de Verdi\`ere Y 2004 {\it Preprint} 

\item[] Colin de Verdi\`ere Y, Lombardi M and Pollet J 1999 {\it
Annales de l'Institut Henri Poincar\'e Physique Theorique} {\bf 71},
95

\item[] Colin de Verdi\'ere Y and Parisse B 1999 {\it
Commun. Math. Phys.} {\bf 205} 459

\item[] Creagh, S C 2004 {\it Nonlinearity} {\bf 17} 1261.

\item[] Dragt A J and Finn J M 1976 {\it J. Math. Phys.} {\bf 17} 2215

\item[] Eckhardt B 1986 {\it J. Phys. A} {\bf 19} 2961

\item[] Emmrich C and Weinstein A 1996 {\it Commun. Math. Phys.} {\bf
176} 701

\item[] Emmrich C and R\"omer H 1988 {\it J. Math. Phys.} {\bf 39} 3530

\item[] Estrada Ricardo, Gracia-Bond\'\i a J M and V\'arilly J C 1989
{\it J. Math. Phys.} {\bf 30} 2789  

\item[] Evans C and Zworski M 2004 {\it Book in preparation}

\item[] Fedosov B V 1994 {\it J. Differential Geometry} {\bf 40} 213

\item[] Gracia-Bond\'\i a J M and V\'arilly J C 1995 {\it
J. Math. Phys.} {\bf 36} 2691

\item[] Gracia-Saz, A 2004 {\it Preprint}.

\item[] Groenewold H J 1946 {\it Physica} {\bf 12} 405

\item[] Hillery M, O'Connell R F, Scully M O and Wigner E P 1984 {\it
Phys. Reports} {\bf 106} 123

\item[] Kontsevich M 2003 {\it Lett. Math. Phys.} {\bf 66} 157

\item[] Kurchan J, Leboeuf P and Saraceno M 1989 {\it Phys. Rev. A} {\bf
40} 6800

\item[] Littlejohn R G 1986 {\it Phys. Reports} {\bf 138} 193

\item[] Littlejohn R G and Flynn W G 1991 {\it Phys. Rev. A} {\bf 44}
5239

\item[]\dash 1992 {\it Chaos} {\bf 2} 149

\item[]\dash 1993 {\it Phys. Rev. Lett.} {\bf 70} 1799

\item[] Littlejohn R G and Weigert Stefan 1993 {\it Phys.\ Rev.\ A}
{\bf 48}, 924

\item[] Maslov V P and Fedoriuk M V 1981 {\it Semi-Classical
Approximations in Quantum Mechanics} (Dordrecht: D. Reidel)

\item[] McDonald S 1988 {\it Phys. Reports} {\bf 158} 377

\item[] Moyal J E 1949 {\it Proc. Camb. Phil. Soc.} {\bf 45} 99

\item[] Sj\"ostrand J and Zworski M 2002 {\it J. Math. Pures Appl.}
{\bf 81} 1

\item[] Voros A 1977 {\it Ann. Inst. Henri Poincar\'e} {\bf 4} 343

\item[]\dash 1989 {\it Phys. Rev. A} {\bf 40} 6814

\item[] Weyl H 1927 {\it Z. Phys.} {\bf 46} 1

\item[] Wigner E P 1932 {\it Phys. Rev.} {\bf 40} 749

\end{harvard}

\end{document}